\documentclass[a4paper,twoside,10pt,english,numbers=noenddot]{scrbook}

\usepackage[english]{babel}
\makeatletter
\def\month@ngerman{\ifcase\month \or Januar\or Februar\or M\"arz\or April\or Mai\or Juni\or Juli\or August\or September\or Oktober\or November\or Dezember\fi}
\def\month@english{\ifcase\month \or January\or February\or March\or April\or May\or June\or July\or August\or September\or October\or November\or December\fi}
\makeatother
\usepackage{hyphenat} 

\usepackage[T1]{fontenc}
\usepackage{helvet} 
\usepackage{indentfirst}
\usepackage[keeplastbox]{flushend} 

\usepackage[usenames,dvipsnames]{xcolor} 
\usepackage{graphicx} 
\usepackage{wrapfig} 
\usepackage{tikz}
\usepackage{rotating}

\usepackage{ifthen}
\usepackage{forloop}
\usepackage{etoolbox}

\usepackage[fleqn]{amsmath} 
\usepackage{amssymb} 
\usepackage{amsfonts} 
\usepackage{amstext} 
\usepackage{wasysym}
\usepackage{mathrsfs} 
\usepackage{mathtools}
\usepackage{upgreek}
\usepackage{siunitx}
\usepackage{esint} 
\usepackage{cancel} 
\usepackage{empheq}
\allowdisplaybreaks

\usepackage{multicol} 
\usepackage{multirow} 
\usepackage{dcolumn}
\usepackage{tabularx} 
\newcolumntype{L}[1]{>{\raggedright\arraybackslash\hsize=#1\hsize}X}
\newcolumntype{R}[1]{>{\raggedleft\arraybackslash\hsize=#1\hsize}X}
\newcolumntype{C}[1]{>{\centering\arraybackslash\hsize=#1\hsize}X}
\usepackage{dblfloatfix} 

\usepackage{enumitem} 
\setlist{nosep} 

\usepackage[a4paper,portrait]{geometry}
\newlength{\TwoColumnWidth}
\newlength{\OneColumnWidth}
\usepackage[headsepline]{scrlayer-scrpage}

\usepackage[hang,bottom]{footmisc}
\deffootnote{\footnotemargin}{0pt}{%
	\textsuperscript{\thefootnotemark}
}
\usepackage[singlelinecheck=false,font=small,labelfont=bf,format=plain]{caption} 
\usepackage[singlelinecheck=true,font=small,labelfont=bf,format=plain]{subfig}
\setkomafont{sectioning}{\rmfamily\bfseries}
\setkomafont{descriptionlabel}{\rmfamily\bfseries}

\usepackage[subfigure]{tocloft} 
\addto\captionsenglish{

}
\makeatletter
  \renewcommand*{\@pnumwidth}{20pt} 
  \renewcommand*{\@tocrmarg}{30pt plus 5pt minus 0pt} 
  \renewcommand*{\@dotsep}{4} 
\makeatother

\newcounter{chapterappendixcounter}[chapter]

\newcounter{totalpagecounter}
\newcounter{totalfigurecounter}
\newcounter{totaltablecounter}
\newcounter{totalcitecounter}
\newcounter{totalpages}
\newcounter{totalfigures}
\newcounter{totaltables}
\newcounter{totalcites}
\makeatletter
\@input{\jobname.myaux}
\makeatother

\definecolor{white}{rgb}{1,1,1}
\definecolor{black}{rgb}{0,0,0}
\definecolor{red}{rgb}{1,0,0}
\definecolor{green}{rgb}{0,1,0}
\definecolor{blue}{rgb}{0,0,1}
\definecolor{cyan}{rgb}{0,1,1}
\definecolor{magenta}{rgb}{1,0,1}
\definecolor{yellow}{rgb}{1,1,0}
\definecolor{darkgreen}{rgb}{0,0.6,0}
\definecolor{darkyellow}{rgb}{0.8,0.8,0}
\definecolor{orange}{rgb}{1,0.5,0}
\definecolor{tuc}{RGB}{0,90,70}
\definecolor{tuclight}{RGB}{218,234,194}
\definecolor{tucorange}{RGB}{242,148,0}
\definecolor{tucbg}{RGB}{224,233,233}

\usepackage[
	colorlinks=true,                   
	urlcolor=blue,                     
	filecolor=blue,                    
	linkcolor=blue,                    
	citecolor=blue,                    
	breaklinks=true,                   
	backref=page,                      
	pagebackref=true,                  
	bookmarks=true,                    
	bookmarksopen=false,               
	bookmarksnumbered=true,            
	pdfauthor={Fabian Teichert},       
	pdfdisplaydoctitle=true,           
	pdfstartview=FitH,                 
	pdfpagelayout=TwoPageRight,        
]{hyperref}

\newif\ifsinglepaper\singlepaperfalse
\addto\captionsenglish{
	\renewcommand\bibname{References}
}

\renewcommand*{\backref}[1]{}
\renewcommand*{\backrefalt}[4]{%
	\ifsinglepaper\else%
		\ifcase #1
		\or (cited at p.~#2).
		\else (cited at pp.~#2).
		\fi%
	\fi
}
\usepackage{cite} 

\newcommand{\bstindent}{99}
\newcommand{\bstaddress}{}
\newcommand{\bstauthor}{}

\newcommand{\bstjournal}{}

\newcommand{\bstpublisher}{}

\newcommand{\bsttitle}{\itshape}
\newcommand{\bstvolume}{}
\newcommand{\bstyear}{}
\newcommand{\bbland}{and}

\newcommand{\bblnov}{November}

\newcommand\bibliographysection{\section}
\newcommand\bibliographysectionstyle{}
\newcommand\bibliographyitemsize{\normalsize}
\newcommand\bibliographyitemseparation{}
\makeatletter
\newcommand\bibcontentsline{\addcontentsline{toc}{section}{References}}
\renewenvironment{thebibliography}[1]{%
	\bibliographysection{\bibliographysectionstyle\bibname}
	\bibcontentsline%
	\renewcommand\rightmark{\bibname}
	\list{\@biblabel{\@arabic\c@enumiv}}{\settowidth\labelwidth{\@biblabel{#1}}%
		\leftmargin\labelwidth%
		\advance\leftmargin\labelsep%
		\@openbib@code%
		\usecounter{enumiv}%
		\let\p@enumiv\@empty%
		\renewcommand\theenumiv{\@arabic\c@enumiv}%
	}%
	\sloppy
	\clubpenalty4000
	\@clubpenalty \clubpenalty
	\widowpenalty4000%
	\sfcode`\.\@m%
}{%
	\def\@noitemerr{\@latex@warning{Empty `thebibliography' environment}}%
	\endlist%
}
\makeatother
\let\oldthebibliography\thebibliography
\renewcommand\thebibliography[1]{
	\bibliographyitemsize
	\oldthebibliography{#1}
	\bibliographyitemseparation
}

\let\oldtwocolumn\twocolumn
\let\oldonecolumn\onecolumn
\newif\iftwocolumn\twocolumntrue
\def\onecolumn{\twocolumnfalse}
\def\twocolumn{\twocolumntrue}
\newif\ifarticlestyle\articlestylefalse

\renewcommand{\title}[2]{%
	\setcounter{authors}{0}%
	\setcounter{addresses}{0}%
	\setcounter{keywords}{0}%
	\def\inserttitle{#1}%
	\def\articlelabel{#2}
}
\newcommand{\email}[1]{\def\insertemail{#1}}
\newcommand{\abstract}[1]{\def\insertabstract{#1}}
\def\insertjournal{}
\def\insertjournalshort{}
\def\insertdoi{}
\def\insertarxiv{}
\def\insertarxivshort{}
\newcommand{\journal}[4][nothing]{%
	\def\insertjournalshort{#2}%
	\if\relax\detokenize{#3}\relax%
		\def\insertjournal{}%
	\else%
		\def\tmpa{#1}%
		\def\tmpb{submitted}%
		\def\tmpc{accepted}%
		\ifx\tmpa\tmpb%
			\def\journalpre{Submitted to: }%
		\else%
			\ifx\tmpa\tmpc%
				\def\journalpre{Accepted in: }%
			\else%
				\def\journalpre{}%
			\fi%
		\fi%
		\if\relax\detokenize{#4}\relax\def\insertjournal{\journalpre#3}\else\def\insertjournal{\journalpre\href{#4}{#3}}\fi%
	\fi%
}
\newcommand{\doi}[1]{\if\relax\detokenize{#1}\relax\def\insertdoi{}\else\def\insertdoi{DOI: \href{http://dx.doi.org/#1}{#1}}\fi}
\newcommand{\arxiv}[2]{\if\relax\detokenize{#1}\relax\def\insertarxiv{}\def\insertarxivshort{}\else\def\insertarxiv{arXiv: \href{https://arxiv.org/abs/#1}{#1 [#2]}}\def\insertarxivshort{arXiv: #1}\fi}
\newcounter{authors}
\newcounter{addresses}
\newcounter{keywords}
\newcommand{\addauthor}[2]{\csdef{author\arabic{authors}}{#1}\csdef{authoraddress\arabic{authors}}{#2}\stepcounter{authors}}
\newcommand{\addaddress}[1]{\csdef{address\arabic{addresses}}{#1}\stepcounter{addresses}}
\newcommand{\addkeyword}[1]{\csdef{keyword\arabic{keywords}}{#1}\stepcounter{keywords}}

\newcounter{otherchapter}
\newcounter{normalchapter}
\newcounter{othercounter}
\newcounter{i}
\newcounter{j}
\makeatletter
\let\normalchapter\chapter

\renewcommand\chapter{%
	\@ifstar{%
		\normalchapter*%
	}{%
		\stepcounter{normalchapter}%
		\normalchapter%
	}%
}
\newcommand\otherchapter{%
	\protected@write\@auxout{}{\string\@writefile{lof}{\string\addvspace{10\string\p@}}}%
	\protected@write\@auxout{}{\string\@writefile{lot}{\string\addvspace{10\string\p@}}}%
	\scr@startsection{chapter}{1}{\z@}{0ex \@plus -0.2ex}{3.5ex \@plus 0.2ex}{\Large\bfseries}%
}
\newcommand\reftype{}
\newcommand\reflabel{}
\newcommand\refnumber{}
\newcommand\refshortnumber{}
\newcommand\reftext{}
\newcommand{\articletitlesub}{%
	\renewcommand\reftype{chapter}%
	\renewcommand\reflabel{section*.\arabic{othercounter}}%
	\renewcommand\refnumber{\Alph{otherchapter}}%
	\renewcommand\refshortnumber{}%
	\renewcommand\reftext{\inserttitle\ifx\insertjournalshort\empty\ (\insertarxivshort)\else\ (\insertjournalshort)\fi}%
	\pdfbookmark[0]{\reftext}{\reflabel}%
	\otherchapter*{\inserttitle}%
	\protected@write\@auxout{}{\string\@writefile{toc}{\string\contentsline {\reftype}{\string\numberline {\refnumber}\reftext}{\thepage}{\reflabel}}}%
	\Alabel{\articlelabel}%
	\noindent\textbf{%
		\large\csuse{author0}$^{\csuse{authoraddress0}}$%
		\forloop{i}{1}{\value{i} < \value{authors}}{%
			, \csuse{author\arabic{i}}$^{\csuse{authoraddress\arabic{i}}}$%
		}
	}\\[1em]
	\normalsize
	\setcounter{j}{0}
	\forloop{i}{0}{\value{i} < \value{addresses}}{%
		\stepcounter{j}
		$^{\arabic{j}}$\,\csuse{address\arabic{i}}
		\ifthenelse{\value{j}<\value{addresses}}{\\}{}
	}
	\ifx\insertemail\empty\\[1em]\else\\[0.5em]E-mail address: \insertemail\\[1em]\fi
	\textbf{Abstract:} \insertabstract
	\ifthenelse{\value{keywords}=0}{}{
		\\[1em]
		Keywords: \csuse{keyword0}%
			\forloop{i}{1}{\value{i} < \value{keywords}}{%
				; \csuse{keyword\arabic{i}}%
			}
	}
}
\newcommand{\articletitle}{%
	\setcounter{articlepage}{0}%
	\stepcounter{otherchapter}%
	\stepcounter{chapter}%
	\setcounter{section}{0}%
	\setcounter{subsection}{0}%
	\setcounter{subsubsection}{0}%
	\stepcounter{othercounter}%
	\iftwocolumn\oldtwocolumn[\articletitlesub\vspace{1.5em}]\else\oldonecolumn\articletitlesub\fi%
}%
\let\oldchapter\chapter
\let\oldsection\section
\let\oldsubsection\subsection
\let\oldsubsubsection\subsubsection
\newcommand\articlesectiondata[1]{%
	\renewcommand\reftype{section}%
	\renewcommand\reflabel{section.\arabic{chapter}.\arabic{section}}%
	\renewcommand\refnumber{\Alph{otherchapter}.\arabic{section}}%
	\renewcommand\refshortnumber{\arabic{section}}%
	\renewcommand\reftext{#1}%
}
\newcommand\articlesubsectiondata[1]{%
	\renewcommand\reftype{subsection}%
	\renewcommand\reflabel{subsection.\arabic{chapter}.\arabic{section}.\arabic{subsection}}%
	\renewcommand\refnumber{\Alph{otherchapter}.\arabic{section}.\arabic{subsection}}%
	\renewcommand\refshortnumber{\arabic{section}.\arabic{subsection}}%
	\renewcommand\reftext{#1}%
}
\newcommand\articlesectionnostar[1]{%
	\articlesectiondata{#1}%
	\pdfbookmark[1]{\reftext}{\reflabel}%
	\scr@startsection{section}{1}{\z@}{-3.5ex \@plus -1ex \@minus -0.2ex}{2.3ex \@plus 0.2ex}{\normalfont\bfseries}{#1}%
	\protected@write\@auxout{}{\string\@writefile{toc}{\string\contentsline {\reftype}{\string\numberline {\refnumber}#1}{\thepage}{\reflabel}}}%
}
\newcommand\articlesectionstar[1]{%
	\articlesectiondata{#1}%
	\scr@startsection{section}{1}{\z@}{-3.5ex \@plus -1ex \@minus -0.2ex}{2.3ex \@plus 0.2ex}{\normalfont\bfseries}*{#1}%
}
\newcommand\articlesubsectionnostar[1]{%
	\articlesubsectiondata{#1}%
	\pdfbookmark[2]{\reftext}{\reflabel}%
	\scr@startsection{subsection}{2}{\z@}{-3.5ex \@plus -1ex \@minus -0.2ex}{2.3ex \@plus 0.2ex}{\normalfont\bfseries}{#1}%
	\protected@write\@auxout{}{\string\@writefile{toc}{\string\contentsline {\reftype}{\string\numberline {\refnumber}#1}{\thepage}{\reflabel}}}%
}
\newcommand\articlesubsectionstar[1]{%
	\articlesubsectiondata{#1}%
	\scr@startsection{subsection}{2}{\z@}{-3.5ex \@plus -1ex \@minus -0.2ex}{2.3ex \@plus 0.2ex}{\normalfont\bfseries}*{#1}%
}
\newcommand\articlesection{\@ifstar{\stepcounter{othercounter}\articlesectionstar}{\articlesectionnostar}}
\newcommand\articlesubsection{\@ifstar{\stepcounter{othercounter}\articlesubsectionstar}{\articlesubsectionnostar}}
\renewcommand\chapter{\@ifstar{\stepcounter{othercounter}\oldchapter*}{\oldchapter}}
\renewcommand\section{\@ifstar{\stepcounter{othercounter}\oldsection*}{\oldsection}}
\renewcommand\subsection{\@ifstar{\stepcounter{othercounter}\oldsubsection*}{\oldsubsection}}
\renewcommand\subsubsection{\@ifstar{\stepcounter{othercounter}\oldsubsubsection*}{\oldsubsubsection}}
\newcommand\listof{}

\newcommand\articlefiguredata{%
	\renewcommand\listof{lof}%
	\renewcommand\reftype{figure}%
	\renewcommand\reflabel{figure.\arabic{chapter}.\arabic{figure}}%
	\renewcommand\refnumber{\Alph{otherchapter}.\arabic{figure}}%
	\renewcommand\refshortnumber{\arabic{figure}}%
}
\newcommand\articletabledata{%
	\renewcommand\listof{lot}%
	\renewcommand\reftype{table}%
	\renewcommand\reflabel{table.\arabic{chapter}.\arabic{table}}%
	\renewcommand\refnumber{\Alph{otherchapter}.\arabic{table}}%
	\renewcommand\refshortnumber{\arabic{table}}%
}
\renewenvironment{figure}{\articlefiguredata\begin{oldfigure}}{\end{oldfigure}} 
\renewenvironment{table}{\articletabledata\begin{oldtable}}{\end{oldtable}} 

\newenvironment{articlefigure*}{\articlefiguredata\begin{figure*}}{\end{figure*}}

\newenvironment{articletable*}{\articletabledata\begin{table*}}{\end{table*}}
\let\oldcaption\caption
\newcommand\Acaption[2][]{%
	\oldcaption[#1]{#2}%
	\renewcommand\reftext{#1}%
	\protected@write\@auxout{}{\string\@writefile{\listof}{\string\contentsline {\reftype}{\string\numberline {\refnumber}#1}{\thepage}{\reflabel}}}%
}
\renewcommand\caption[2][]{\ifarticlestyle\Acaption[#1]{#2}\else\oldcaption[#1]{#2}\fi}

\newcommand\botholdlabel[1]{\oldlabel{#1}\oldlabel{A#1}}
\newenvironment{articleequation}{\begin{equation}\renewcommand\label{\botholdlabel}}{\end{equation}} 
\newcommand\Aref[1]{\oldref{A#1}}
\newcommand\Alabel[1]{%
	\protected@write\@auxout{}{\string\newlabel{#1}{{\refnumber}{\thepage}{\reftext}{\reflabel}{}}}%
	\protected@write\@auxout{}{\string\newlabel{A#1}{{\refshortnumber}{\thepage}{\reftext}{\reflabel}{}}}%
}
\AtBeginDocument{%
	\let\oldref\ref%
	\let\oldlabel\label%
	\renewcommand\ref[1]{\ifarticlestyle\Aref{#1}\else\oldref{#1}\fi}%
	\renewcommand\label[1]{\ifarticlestyle\Alabel{#1}\else\oldlabel{#1}\fi}%
}
\makeatother

\newcommand\articlestyleheaderleft{%
	\ifx\insertarxiv\empty%
		\ifx\insertjournal\empty\linebreak\textnormal\insertdoi\else\linebreak\textnormal\insertjournal\fi%
	\else%
		\ifx\insertdoi\empty\linebreak\textnormal\insertjournal\else\textnormal\insertjournal\linebreak\textnormal\insertdoi\fi%
	\fi%
}
\newcommand\articlestyleheaderright{%
	\ifx\insertarxiv\empty%
		\ifx\insertjournal\empty\else\linebreak\textnormal\insertdoi\fi%
	\else%
		\linebreak\textnormal\insertarxiv%
	\fi%
}
\newcommand\articlestyleheadercenter{%
	\linebreak\textnormal\thechapter
}

\newcounter{articlepage}
\newcommand\articlepagemark{\arabic{articlepage}}

\newcommand\nocontentsline[3]{}
\let\oldaddcontentsline\addcontentsline
\newcommand\normalstyle{%
	\articlestylefalse%
	\KOMAoptions{fontsize=11pt}%
	\newgeometry{left=3cm,right=2.5cm,top=4cm,bottom=4cm}
	\setlength{\headheight}{26pt}
	\setlength{\headsep}{24pt}
	\setlength{\footskip}{30pt}
	\setlength{\TwoColumnWidth}{\textwidth}
	\setlength{\OneColumnWidth}{0.5\TwoColumnWidth-0.5\columnsep}
	\clearpairofpagestyles%
	\ihead{}%
	\chead{}%
	\ohead{\ifthispageodd{\textnormal\rightmark}{\textnormal\leftmark}}%
	\ifoot{}%
	\cfoot{}%
	\ofoot[\textnormal\pagemark]{\textnormal\pagemark}%
	\let\addcontentsline\oldaddcontentsline%
	\renewcommand{\thechapter}{\arabic{normalchapter}}%
	\renewcommand{\thesection}{\arabic{normalchapter}.\arabic{section}}%
	\renewcommand{\thesubsection}{\arabic{normalchapter}.\arabic{section}.\arabic{subsection}}%
	\renewcommand{\thesubsubsection}{\arabic{normalchapter}.\arabic{section}.\arabic{subsection}.\arabic{subsubsection}}%
	\renewcommand{\thefigure}{\arabic{normalchapter}.\arabic{figure}}%
	\renewcommand{\thetable}{\arabic{normalchapter}.\arabic{table}}%
	\renewcommand{\theequation}{\arabic{normalchapter}.\arabic{equation}}%
	\renewcommand\bibliographysection{\section*}%
	\renewcommand\bibcontentsline{\addcontentsline{toc}{section}{References}}
	\renewcommand\bibliographysectionstyle{}%
	\renewcommand\bibliographyitemsize{\normalsize}%
	\renewcommand\bibliographyitemseparation{%
		\setlength{\parskip}{0pt}%
		\setlength{\itemsep}{5pt plus 0.3ex}%
	}%
	\allowdisplaybreaks%
}
\newcommand\articlestyle{%
	\articlestyletrue%
	\KOMAoptions{fontsize=10pt}%
	\newgeometry{left=1.5cm,right=1.5cm,top=2.95cm,bottom=1.55cm}
	\setlength{\headheight}{24pt}
	\setlength{\headsep}{20pt}
	\setlength{\footskip}{1.2cm}
	\setlength{\TwoColumnWidth}{\textwidth}
	\setlength{\OneColumnWidth}{0.5\TwoColumnWidth-0.5\columnsep}
	\clearpairofpagestyles%
	\ihead{\ifthispageodd{\articlestyleheaderleft}{\articlestyleheaderright}}%
	\chead{\ifsinglepaper\else\articlestyleheadercenter\fi}%
	\ohead{\ifthispageodd{\articlestyleheaderright}{\articlestyleheaderleft}}%
	\ifoot{}%
	\cfoot{\ifsinglepaper\textnormal\pagemark\else\stepcounter{articlepage}\textnormal{\thechapter-\articlepagemark}\fi}%
	\ofoot{\ifsinglepaper\else\textnormal\pagemark\fi}%
	\let\addcontentsline\nocontentsline%
	\renewcommand{\thechapter}{\Alph{otherchapter}}%
	\renewcommand{\thesection}{\arabic{section}}%
	\renewcommand{\thesubsection}{\arabic{section}.\arabic{subsection}}%
	\renewcommand{\thesubsubsection}{\arabic{section}.\arabic{subsection}.\arabic{subsubsection}}%
	\renewcommand{\thefigure}{\arabic{figure}}%
	\renewcommand{\thetable}{\arabic{table}}%
	\renewcommand{\theequation}{\arabic{equation}}%
	\renewcommand\bibliographysection{\articlesection*}%
	\renewcommand\bibcontentsline{\oldaddcontentsline{toc}{section}{References}}
	\renewcommand\bibliographysectionstyle{\normalsize}%
	\renewcommand\bibliographyitemsize{\small}%
	\renewcommand\bibliographyitemseparation{%
		\setlength{\parskip}{0pt}%
		\setlength{\itemsep}{0pt plus 0.3ex}%
	}%
	\interdisplaylinepenalty=10000%
}
\normalstyle

\renewcommand{\hbar}{\mathchar'26\mkern-9mu \mathrm{h}}
\newcommand{\one}{\mathcal{I}}
\newcommand{\hamilton}{\mathcal{H}}

\newcommand{\coupling}{\tau}

\newcommand{\green}{\mathcal{G}}

\newcommand{\transmission}{\mathcal{T}}

\newcommand{\imag}{\text{i}}

\newcommand{\trace}[1]{\text{Tr}\!\left[#1\right]} 
\newcommand{\order}[1]{\mathcal{O}\!\left(#1\right)}

\newcommand{\sref}[1]{section \ref{#1}}

\newcommand{\fref}[1]{figure \ref{#1}}
\newcommand{\Fref}[1]{Figure \ref{#1}}

\newcommand{\eref}[1]{(\ref{#1})}

\singlepapertrue
\begin{document}

\raggedbottom

\frontmatter
\clearpairofpagestyles
\ofoot[\textnormal\pagemark]{\textnormal\pagemark}
\KOMAoptions{headsepline=false}

\mainmatter
\KOMAoptions{headsepline=true}

\normalstyle
\articlestyle

\renewcommand{\one}{\mathcal{I}}
\renewcommand{\hamilton}{\mathcal{H}}
\renewcommand{\coupling}{\tau}
\renewcommand{\green}{\mathcal{G}}
\renewcommand{\transmission}{\mathcal{T}}
\renewcommand{\imag}{\text{i}}
\renewcommand{\order}[1]{\mathcal{O}\!\left(#1\right)}
\newcommand{\avg}[1]{\left\langle#1\right\rangle} 
\renewcommand{\trace}[1]{\text{Tr}\!\left[#1\right]} 
\def\intd#1#2#3#4{\int\limits_{#1}^{#2}#3\mathrm{d}#4} 

\onecolumn 

\title{Strong localization in defective carbon nanotubes: a recursive Green's function study}{NJP}

\addauthor{Fabian Teichert}{1,2}
\addauthor{Andreas Zienert}{1}
\addauthor{J\"org Schuster}{3}
\addauthor{Michael Schreiber}{2}

\addaddress{Institute of Physics, Technische Universit\"at Chemnitz, 09107 Chemnitz, Germany}
\addaddress{Center for Microtechnologies, Technische Universit\"at Chemnitz, 09107 Chemnitz, Germany}
\addaddress{Fraunhofer Institute for Electronic Nano Systems (ENAS), 09126 Chemnitz, Germany}
\addaddress{Dresden Center for Computational Materials Science (DCMS), TU Dresden, 01062 Dresden, Germany}

\email{fabian.teichert@physik.tu-chemnitz.de}

\abstract{We study the transport properties of defective single-walled armchair carbon nanotubes (CNTs) on a mesoscopic length scale. Monovacancies and divancancies are positioned randomly along the CNT. The calculations are based on a fast, linearly scaling recursive Green's function formalism that allows us to treat large systems quantum-mechanically. The electronic structure of the CNT is described by a density-functional-based tight-binding model. We determine the influence of the defects on the transmission function for a given defect density by statistical analysis. We show that the system is in the regime of strong localization (e.g. Anderson localization). In the limit of large disorder the conductance scales exponentially with the number of defects. This allows us to extract the localization length. Furthermore, we study in a systematic and comprehensive way, how the conductance, the conductance distribution, and the localization length depend on defect probability, CNT diameter, and temperature.}

\addkeyword{carbon nanotube (CNT)}
\addkeyword{defect}
\addkeyword{density-functional-based tight binding (DFTB)}
\addkeyword{electronic transport}
\addkeyword{recursive Green's function formalism (RGF)}
\addkeyword{strong localization}

\journal{New J. Phys. 16 (2014), 123026}{New Journal of Physics 16 (2014), 123026}{http://iopscience.iop.org/article/10.1088/1367-2630/16/12/123026} 
\doi{10.1088/1367-2630/16/12/123026} 
\arxiv{1705.01757}{cond-mat.mes-hall} 

\articletitle

\articlesection{Introduction}\label{NJP:Introduction}

Carbon nanotubes (CNTs) \cite{Nature.354.56} offer remarkable electronic properties \cite{RevModPhys.79.677, NanoRes.1.361} and thus, they are of great interest for different applications in nanotechnology. Semiconducting CNTs are applied for example as a channel in field effect transistors \cite{Nature.393.49} and metallic ones are used as interconnects \cite{MicroEng.64.399, Nature.393.240}. Like in any other technology, several challenges have to be overcome.

One challenge is the unavoidable presence of defects even in CNTs of the highest quality. Mono- and divacancies are the most common intrinsic defects. They can be created during irradiation \cite{PhysRevB.63.245405, NanoLett.9.2285} and plasma processes \cite{JPhysDApplPhys.43.305402}, which are necessary for nanoelectronic applications. Furthermore, one can find vacancy clusters, crystallographic defects, substitutional defects, impurities, and functionalizations. CNTs which are shorter than the mean free path of the electrons show ballistic conduction. That means the absence of scattering processes, leading to the highest possible conductance. The introduction of defects has a strong influence on the conductivity \cite{Science.272.523, Nature.382.54}. The ballistic transport in clean CNTs is driven into the regime of strong localization \cite{PhysRevB.64.045409}. This has been studied selectively for vacancies \cite{JPhysCondMat.20.294214, JPhysCondMat.20.304211, PhysRevLett.95.266801}, vacancy clusters \cite{JPhysChemC.116.1179}, substitutional atoms \cite{SolidStateCommun.149.874}, and functionalizations \cite{PhysStatSolB.247.2962, NanoRes.3.288, NanoLett.9.940}. Experimental studies confirm the theoretically predicted localization regime \cite{NatureMaterials.4.534}. The results are qualitatively similar to CNTs with disorder described by the Anderson model of localization \cite{PhysRevB.58.4882}.

Due to the lack of devices  based on high quality CNTs with well defined defect properties a systematic experimental study of the influence of defects on the electron transport properties is still out of reach. Thus, numerical calculations with efficient algorithms are the method of choice to derive general dependencies.

To calculate properties of quantum mechanical systems, density functional theory (DFT) is a sufficiently exact and common tool to deal with systems of up to thousand atoms. However, the typical $\order{N^3}$ scaling (where $N$ is the number of atoms) of the computational effort, excludes DFT from applications in the mesoscopic range with several hundred thousand atoms. Although the search for linear scaling DFT approaches is a very active field of research \cite{RepProgPhys.75.036503}, standard DFT codes can not be applied to the large systems we want to describe.

We avoid these limitations by using the recursive (equilibrium and non-self-consistent) Green's function formalism (RGF) for our transport calculations \cite{JPhysCSolidStatePhys.14.235}, described in \sref{NJP:RGF}. The RGF is a very efficient and fast algorithm for one-dimensional systems with short-range interactions. Its computational effort scales linearly with the number of atoms in the system.

To describe the underlying electronic structure, we use the density-functional-based tight-binding (DFTB) approach \cite{PhysRevB.51.12947, IntJQuantumChem.58.185}, which is a hybrid of DFT and TB. It combines the accuracy of DFT with the computational simplicity of TB and therefore, it is a perfect model to be used as basis for the RGF.

The present paper describes a comprehensive and systematic study of transport properties of defective armchair CNTs. We focus on mono- and divacancies, which are the most common defects occurring in CNTs. We calculate the transmission function and the conductivity of random defect configurations for various defect probabilities and tube lengths, and quantify the scaling behaviour of the mean conductance by extracting the localization length. Furthermore, the influence of the temperature on the statistical variation of the conductance of individual CNTs as well as on the mean conductance and the localization length is discussed. We determine the dependence of the localization length on the tube diameter for different defect types and temperatures for the first time in a systematic way with a consistent theory.

\articlesection{Theoretical framework}\label{NJP:Theory}

To describe the electronic transport of defective CNTs we use a fast, linearly scaling RGF, which is embedded in the common equilibrium Green's function transport formalism~\cite{Datta2005}.

\articlesubsection{Transport formalism}\label{NJP:Transport}

\begin{figure}
	\includegraphics{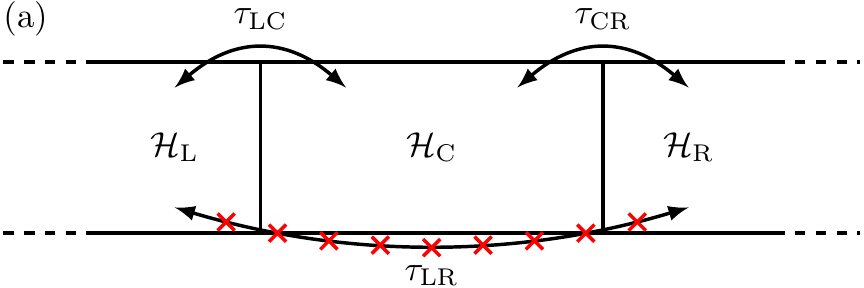}\hfill
	\includegraphics{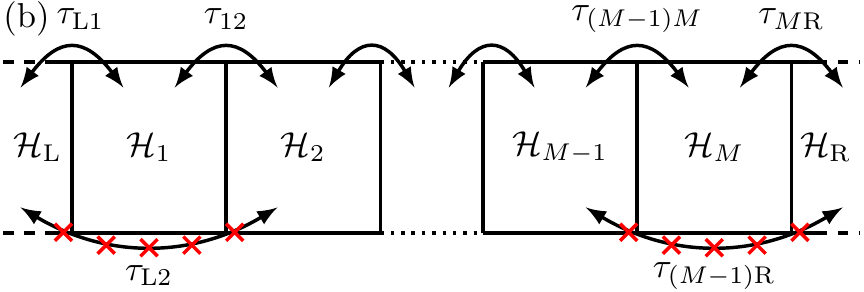}
	\caption[Scheme of a device system]{Scheme of the whole system used in the transport formalism. The Hamilton matrices of the subsystems are denoted by $\hamilton$ and the coupling matrices between the subsystems by $\coupling$. (a) The two electrodes (L and R) consist of ideal, semi-infinite CNTs. The central region (C) includes the defective CNT. (b) The central region is divided into $N$ subsystems to which the RGF is applied.}
	\label{NJP:fig:transport:scheme}
\end{figure}

We consider an infinite CNT with a finite number of defects. To handle this, the whole system is treated as a device configuration, which is sketched in \fref{NJP:fig:transport:scheme}a. The device consists of two semi-infinite electrodes and a finite central region, which contains the defects. Such a partition is due to the requirement of numerical simplicity, because the treatment of the whole infinite and non-periodic system is computationally impracticable. By separating the electrodes from the defective central region, they can be handled like a bulk system, using the very efficient iterative renormalization decimation algorithm (RDA)~\cite{JPhysFMetPhys.14.1205, JPhysFMetPhys.15.851}. Thus, the Schr\"odinger equation of the infinite system can be reduced to an effective Schr\"odinger equation for the finite central region.

The Schr\"odinger equation of the whole device within an orthonormal basis reads
\begin{articleequation}
	\left(\begin{array}{ccc}
		\hamilton_\text{L} & \coupling_\text{LC} & 0 \\
		\coupling_\text{CL} & \hamilton_\text{C} & \coupling_\text{CR} \\
		0 & \coupling_\text{RC} & \hamilton_\text{R}
	\end{array}\right)\varPsi = E\varPsi \label{NJP:eqn:SGL}
\end{articleequation}%
with Hamilton matrices $\hamilton$ and coupling matrices $\coupling$, corresponding to the parts L, R, and C. There is no coupling between the two electrodes, because this can be neglected by choosing the central region always larger than the interaction range of its atoms. For a non-orthogonal basis with an additional overlap matrix the system can be transformed into a form like \eref{NJP:eqn:SGL}. The subsequent equations can easily be obtained in this case by substituting the corresponding Hamilton matrices and coupling matrices.

We define the advanced Green's function matrix of the central region
\begin{articleequation}
	\green_\text{C}\!\left(E\right) = \lim_{\eta\rightarrow0^+} \left[ (E+\imag\eta)\one - \hamilton_\text{C} - \varSigma_\text{L} - \varSigma_\text{R} \right]^{-1} \qquad .\label{NJP:eqn:Green}
\end{articleequation}%
The selfenergy matrices $\varSigma_\text{L} = \coupling_\text{CL} \green_\text{L} \coupling_\text{LC}$ and $\varSigma_\text{R} = \coupling_\text{CR} \green_\text{R} \coupling_\text{RC}$ describe the coupling to the electrodes and lead to an energy shift of the related states. The advanced surface Green's functions of the electrodes $\green_\text{L/R}$ are calculated by the RDA. $\one$ is an identity matrix of required dimension. The mathematical limit $\eta\rightarrow 0^+$ is not feasible. In fact, $\eta$ will be used as a numerical parameter, which is set to a sufficiently small, fixed value to get well conditioned matrices and small errors for the inversion \eref{NJP:eqn:Green} near eigenvalues and a good convergence of the RDA. We choose $\eta=10^{-4}$ for calculating the Fermi energy with RDA.

The transmission spectrum $\transmission(E)$ of the device configuration, which is given by the sum over all transmission probabilities of accessible transmission channels, is calculated using
\begin{articleequation}
	\transmission(E) = \trace{\varGamma_\text{R} \green_\text{C} \varGamma_\text{L} \green_\text{C}^\dagger} \qquad .\label{NJP:eqn:T(E)}
\end{articleequation}%
The matrices $\varGamma_\text{L/R} = \imag ( \varSigma_\text{L/R} - \varSigma_\text{L/R}^\dagger )$ describe the broadening of the energy levels of the central region due to the coupling to the electrodes.

The conductance $G$ in the limit of vanishing applied voltage is calculated within the Landauer-B\"uttiker formalism~\cite{PhysRevB.31.6207}. Thus, it is determined as
\begin{articleequation}
	G = -\text{G}_0\intd{-\infty}{\infty}{\transmission(E)f'(E)}{E} \qquad ,\label{NJP:eqn:G}
\end{articleequation}%
where $f(E)$ is the Fermi distribution function and $\text{G}_0=2\text{e}^2/\text{h}\approx(\SI{12.9}{\kilo\ohm})^{-1}$ is the conductance quantum. The temperature dependence of $G$ is contained in $f(E)$. At zero temperature it follows
\begin{articleequation}
	G = \text{G}_0\transmission(E_\text{F}) \qquad ,\label{NJP:eqn:G:0K}
\end{articleequation}%
where $E_\text{F}$ is the Fermi energy.

Note that only elastic scattering at static defects is described by the present theory. Phonons are neglected. Thus, the results are limited to the coherent regime. The coherent regime is either achieved by small tube lengths (smaller than the phase coherence length) or attained by large defect densities (which causes elastic scattering to dominate). Otherwise, decoherence effects must be taken into account.

The neglection of decoherence effects in our work is justified by the fact that the defective regions are up to $\SI{2460}{\nano\metre}$ long. Acoustic phonons have a coherence length of $\SI{2400}{\nano\metre}$ \cite{NanoLett.4.517}, whereas optical phonons have a coherence length in the range of ten to hundred $\si{\nano\metre}$ \cite{NanoLett.4.517, PhysRevLett.92.106804}. But in contrast to acoustic phonons, optical phonons have energies above thermal fluctuations and can be neglected in the low bias regime. Consequently, the tube lengths used here are small enough to assume coherent transport. In \cite{PhysRevLett.104.116801} it is shown that a (5,5)-CNT with Anderson-type disorder of certain strength has a characteristic temperature $T_\text{c}=\SI{250}{\kelvin}$, above which phonon scattering dominates. The authors get an elastic scattering length of about $\SI{690}{\nano\metre}$. The defect probabilities that we use (see \sref{NJP:Model}), lead to a lower elastic scattering length, i.e. higher $T_\text{c}$. For example, the temperature $T_\text{c}$ is above $T=\SI{500}{\kelvin}$ for defect probabilities $p_\text{D}>\num{0.001}$\footnote{For this comparison, we made the simple assumption that the elastic scattering length is equal to the mean distance of the center of neighbouring defects. Then, we get the condition $\frac{1}{2}\cdot\SI{690}{\nano\metre}>\SI{0.246}{\nano\metre}/p_\text{D}$, where the factor $\frac{1}{2}$ denotes the reduction of the elastic scattering length at $T=\SI{500}{\kelvin}$ in comparison to $T=\SI{300}{\kelvin}$ (see inset of figure 3 in \cite{PhysRevLett.104.116801}).}.

\articlesubsection{Recursive Green's function formalism}\label{NJP:RGF}

The calculation time of the formalism shown so far scales as $t\sim\order{[\dim\hamilton_\text{C}]^3}$, because the inversion \eref{NJP:eqn:Green} is the most time-consuming process. Calculating very large systems with several hundred thousand atoms would not be feasible with respect to calculation times and memory requirements. However, the RGF \cite{JPhysCSolidStatePhys.14.235} provides an algorithm which avoids these problems and is especially well suited for quasi one-dimensional systems with short-range interaction.

We consider a device configuration with a central region, whose length is much larger than the interaction range of an atom. Thus, it can be divided into $N$ subsystems in such a way, that every cell interacts only with its nearest-neighbour cells, as shown in \fref{NJP:fig:transport:scheme}b. Thus, we neglect the coupling between non-nearest-neighbour cells in the same way as we neglect the coupling between the two electrodes in \fref{NJP:fig:transport:scheme}a. It follows, that $\hamilton_\text{C}$ is a block-tridiagonal matrix with $N\!\times\!N$ blocks. In this case, the transmission spectrum reads
\begin{articleequation}
	\transmission(E) = \trace{\varGamma'_\text{R} \green_{N1} \varGamma'_\text{L} \green_{N1}^\dagger} \qquad ,\label{NJP:eqn:T(E):RGF}
\end{articleequation}%
where $\green_{N1}$ is the lower left block of the full Green's function matrix $\green_\text{C} = \left\{\green_{ij}\right\}_{i,j=1}^{N}$ and $\varGamma'_\text{L/R}$ is the upper left respectively lower right block of the full broadening matrix $\varGamma_\text{L/R}$.

Applying \eref{NJP:eqn:T(E):RGF} we gain a factor $N$ in calculation time and required memory compared to \eref{NJP:eqn:T(E)}, because calculating all $\green_{ij}$ is more efficient than calculation $\green_\text{C}$ with direct inversion. We gain another factor $N$, because only the matrix block $\green_{N1}$ is needed. This is possible since $\green_{N1}$ can be calculated directly without knowledge of all other matrix blocks. In order to get further insight into the sequence of RGF steps when regarding a scattering problem the equations can be written as
\begin{articleequation}
	\begin{aligned}
		\green_{i} &= \lim_{\eta\rightarrow0^+} \left[ (E+\imag\eta)\one - \hamilton_\text{i} \right]^{-1} \qquad ,\\
		\mathcal{S}_1 &= \one \qquad ,\\
		\mathcal{S}_i &= \left( \one-\green_i\coupling_{i(i-1)}\mathcal{S}_{i-1}\green_{i-1}\coupling_{(i-1)i} \right)^{-1} \qquad ,\\
		\mathcal{P}_{i(i-1)} &= \mathcal{S}_i\green_i\coupling_{i(i-1)} \qquad ,\\
		\green_{N1} &= \mathcal{P}_{N(N-1)}\mathcal{P}_{(N-1)(N-2)}\cdots\mathcal{P}_{32}\mathcal{P}_{21}\green_1 \qquad .
	\end{aligned}\label{NJP:eqn:RGF}
\end{articleequation}
Therein, $\green_i$ are the Green's functions of the isolated subsystems (see \fref{NJP:fig:transport:scheme}). $\mathcal{S}_i$ are multiple scattering factors for scattering from the first to the $i$th cell. $\mathcal{P}_{i(i-1)}$ are forward propagation factors for the propagation of a particle from the $(i-1)$-th to the $i$th cell. The calculation of the transmission with RGF is done with a smaller value of the numerical parameter $\eta=10^{-7}$ due to the larger number of matrix multiplications compared to the Fermi-energy computation with RDA.

In summary, this formalism shows a scaling behaviour as $\order{N[\dim\hamilton_\text{i}]^3}$. For fixed cells $\hamilton_i$ of similar size we get a calculation time which scales linear with $N$, where $N$ is a measure of the CNT length. This is a gigantic improvement of the original formalism and thus makes transport properties of larger mesoscopic systems accessible.

\articlesubsection{Electronic structure}\label{NJP:ElectronicStructure}

We describe the electronic structure of each cell by using a DFTB model~\cite{PhysRevB.51.12947, IntJQuantumChem.58.185}, taking advantage of both methods: the high speed of the TB approach for creating the Hamilton matrices and overlap matrices, and the accuracy and transferability of DFT calculations. The DFTB model allows us to create the required matrices during the algorithm instantaneously without storing them, what reduces the required memory. Furthermore, coupling between different ideal and arbitrary defective cells can be included, which would be quite difficult using plain DFT. We also looked at a more simple orthogonal TB approach with nearest-neighbour hopping but this can not reproduce the transmission spectra of defective CNTs sufficiently well, compared to DFT calculations.

For our calculations we use the existing DFTB parameter set 3ob~\cite{JChemTheoryComput.9.338}, which is available at www.dftb.org. It utilizes a non-orthogonal sp$^3$ basis and provides the TB parameters for carbon, hydrogen, oxygen, and nitrogen. It is an extension and improvement of the former parameter set mio~\cite{PhysRevB.58.7260}. The TB parameters were extracted from DFT Hamiltonians and reproduce bond lengths, bond angles, binding energies, and vibrational frequencies of selected molecules like water, ammonia, and in particular benzene. Besides the sets 3ob and mio, we checked three other parameter sets, developed by Hancock \cite{JLowTempPhys.153.393}, Vogl \cite{JPhysChemSolids.44.365}, and Jancu \cite{PhysRevB.57.6493}. The CNT bandstructure obtained with 3ob was found to show the best agreement with the DFT results in an energy range of $\pm\SI{1}{\electronvolt}$ around the Fermi energy.

The carbon hopping and overlap parameters of the 3ob set vanish rapidly with increasing atom distances. For choosing a cutoff radius, two things have to be considered: Firstly, a larger cutoff radius leads to smaller errors within the DFBT model. Secondly, a large cutoff radius requires larger cells due to the restriction of the RGF that cells must not interact beyond nearest neighbours. As a good compromise we fix the cutoff radius at twice the carbon distance in the nanotube. Beyond that distance the parameters are sufficiently small. Considering the ideal CNT, this leads to a third-nearest-neighbour description. Thus, the resulting ideal unit cell of the algorithm consists of only one primitive CNT cell, which is as small as possible (length $a_\text{UC}\approx \SI{2.46}{\angstrom}$).

The DFTB Hamiltonians are obtained non-self-consistently and the present RGF formalism is only suitable for transport at equilibrium. However, as we focus on the limit of small bias voltages in absence of electromagnetic fields, these approximations are reasonable.

\articlesubsection{Strong localization}\label{NJP:Localization}

Mesoscopic transport is governed by different microscopic mechanisms, which compete with each other and influence the electronic properties. Thus, several transport regimes exist within certain characteristic lengths. In the previously described formalism, only elastic scattering is considered, which leads to ballistic transport for CNTs shorter than the elastic scattering length and diffusive transport for longer CNTs. Furthermore, the electronic states of a quantum mechanical system can become localized, e.g. due to interference effects. Often, their wave functions decay exponentially in real space. These states are called exponentially localized states and the corresponding decay length is labeled localization length. Systems larger than the localization length are driven from the diffusive transport regime into the strong localization regime, where the conductance decreases exponentially with the system size.

For the Anderson model of localization it can be shown that with increasing disorder an increasing number of states is localized, yielding a metal-insulator transition in disordered bulk systems \cite{PhysRev.109.1492, Mello2004}. Especially in one-dimensional systems, like the ones considered here, arbitrarily small disorder already leads to the localized regime \cite{PhysRevLett.47.1546, PhysRevLett.42.673}. It is reasonable to expect that this will also be valid for randomly distributed realistic defects instead of the commonly investigated energetic disorder in terms of random diagonal matrix elements of $\hamilton$.

\begin{figure}[tb]
	\begin{minipage}{0.0667\textwidth}
		\centering
		$\text{UC}$
		\includegraphics[width=\textwidth,height=2.4\textwidth]{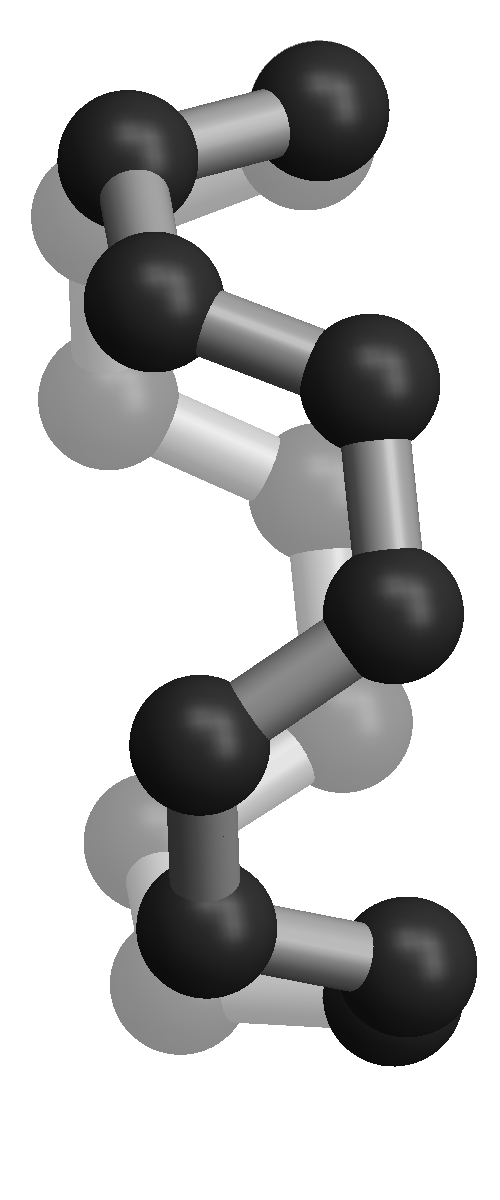}
	\end{minipage}\hfill
	\begin{minipage}{0.0667\textwidth}
		\centering
		$\text{MV}$
		\includegraphics[width=\textwidth]{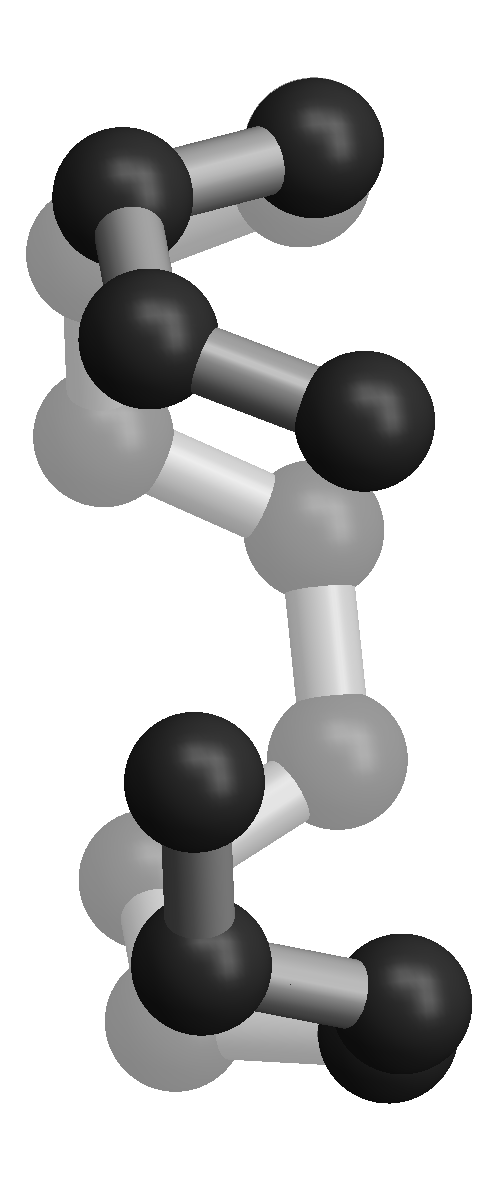}
	\end{minipage}\hfill
	\begin{minipage}{0.2\textwidth}
		\centering
		$\text{MV}_\text{3H}$
		\includegraphics[width=\textwidth]{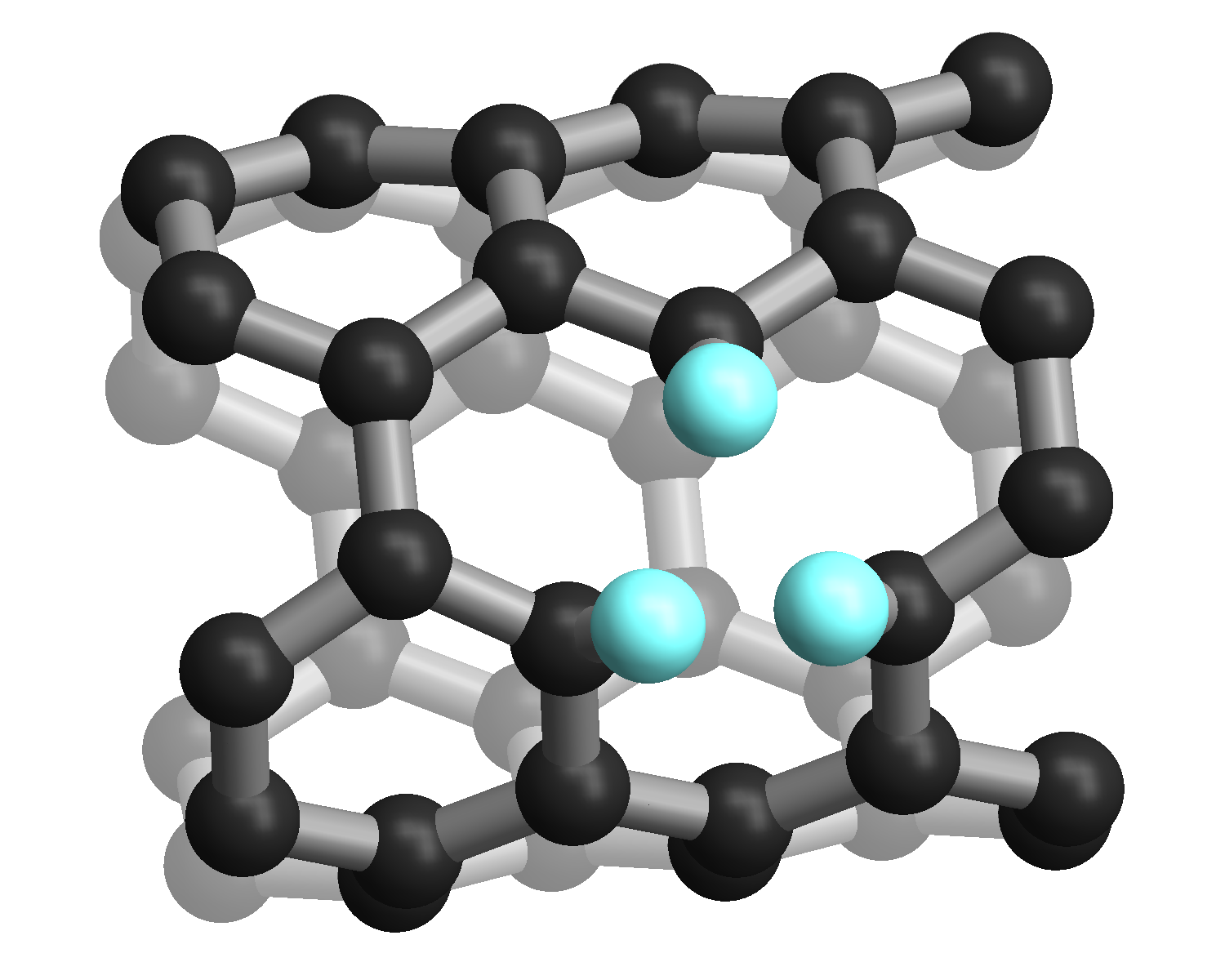}
	\end{minipage}\hfill
	\begin{minipage}{0.2\textwidth}
		\centering
		$\text{DV}_\text{perp}$
		\includegraphics[width=\textwidth]{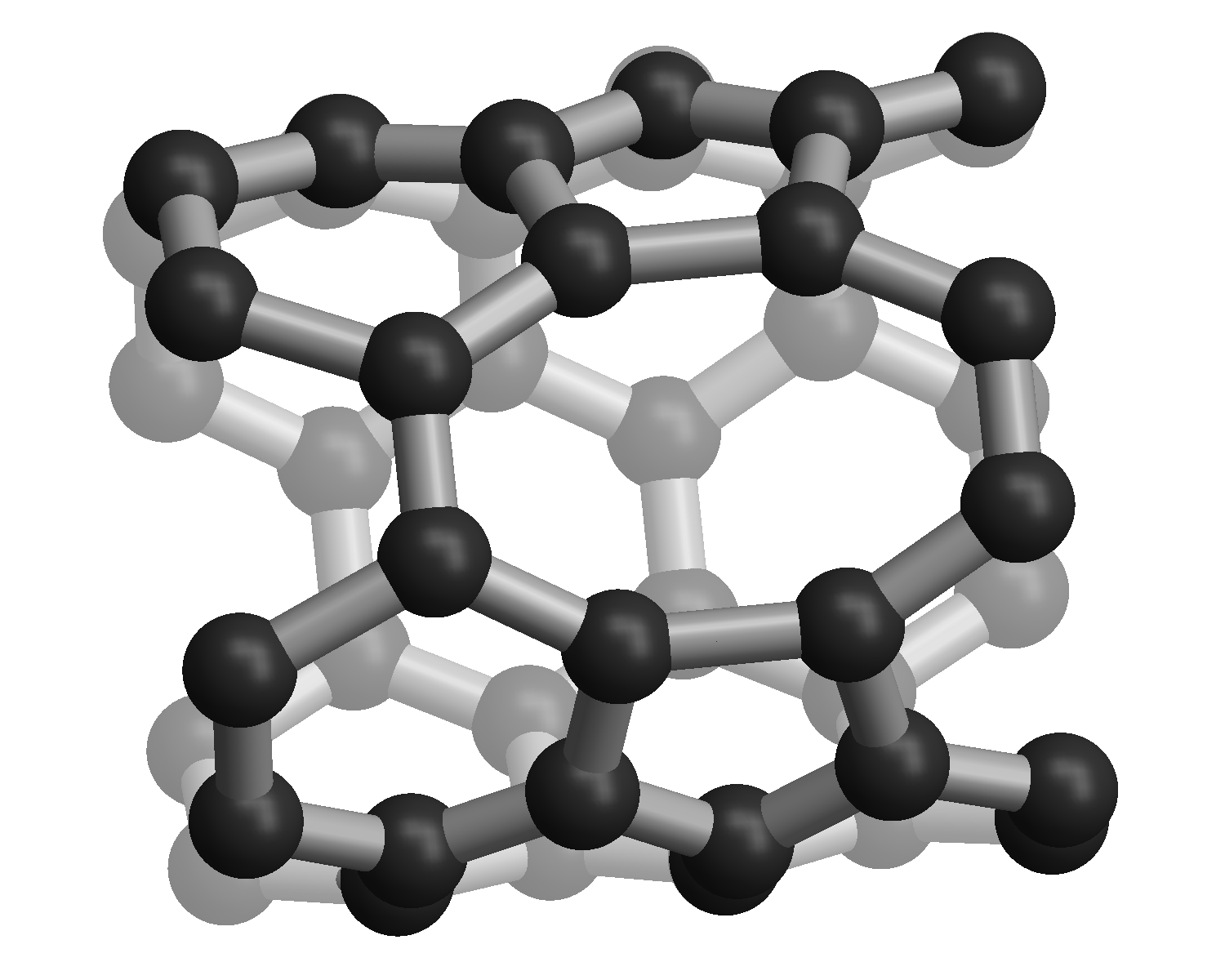}
	\end{minipage}\hfill
	\begin{minipage}{0.2\textwidth}
		\centering
		$\text{DV}_\text{diag}$
		\includegraphics[width=\textwidth]{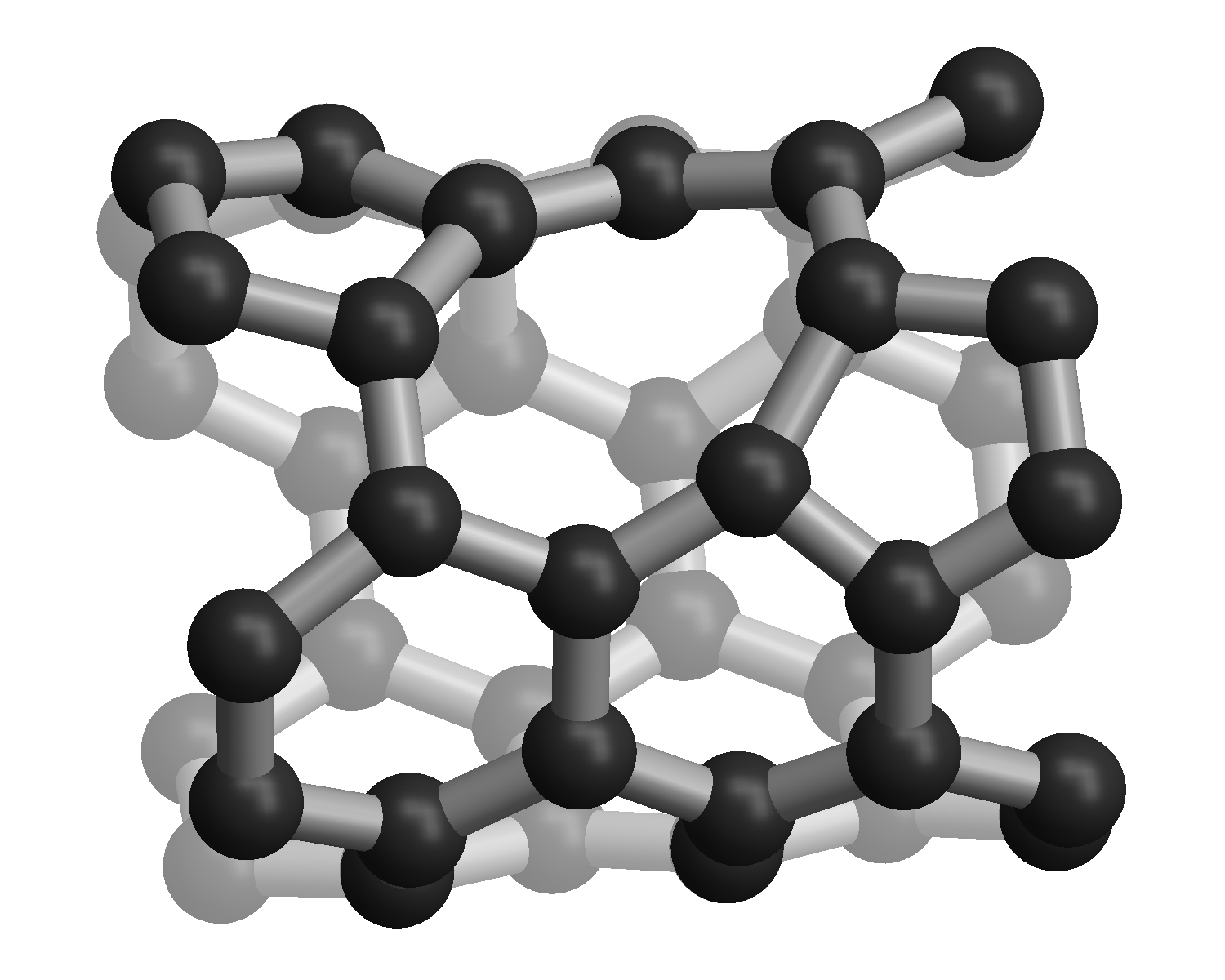}
	\end{minipage}\hfill
	\begin{minipage}{0.24\textwidth}
		\centering
		$\text{DV}_\text{diag}$
		\includegraphics[width=\textwidth]{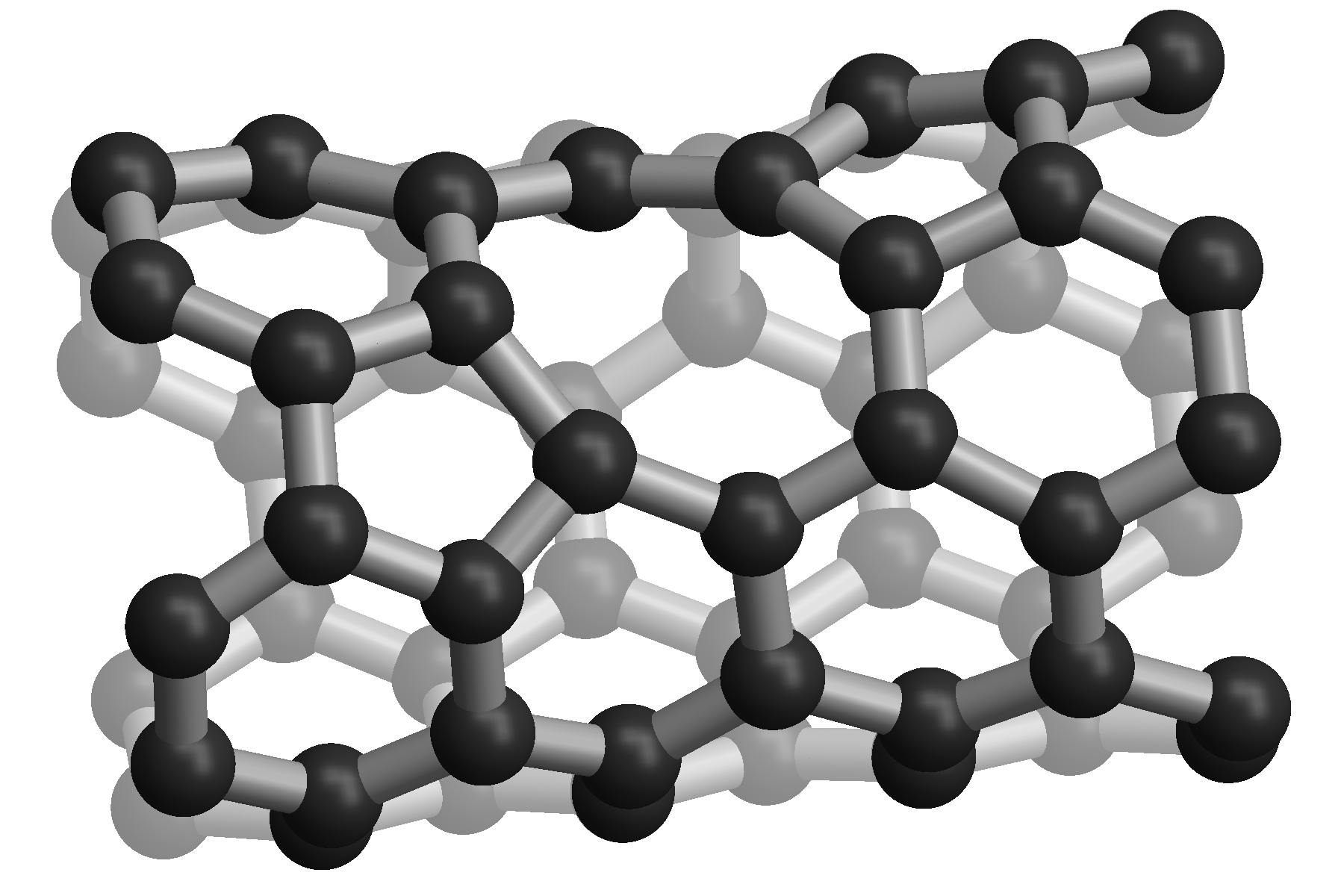}
	\end{minipage}
	\caption[Geometric structures of the CNT cells]{Geometry of the used cells, exemplarily shown for the (4,4)-CNT: unit cell of the ideal CNT (UC), unsaturated monovacancy (MV), saturated monovacancy (MV$_\text{3H}$), divacancy with perpendicular orientation (DV$_\text{perp}$), and the two types of the divacancy with diagonal orientation (DV$_\text{diag}$). Each subsystem for the RGF (\fref{NJP:fig:transport:scheme}b) is chosen as one of these cells. For larger tubes see supplementary data.}
	\label{NJP:fig:defect:geometry}
\end{figure}

\articlesection{Modeling details of the defective system}\label{NJP:Model}

The aim of the present work is to describe the influence of realistic defects on the electronic transport properties of long metallic CNTs. In general, the geometry of a ($m$,$n$)-CNT is characterized by two integers $m$ and $n$, which are called chiral indices. $m$ and $n$ define a rectangular segment in the graphene sheet, which is rolled up to build the CNT. The diameter of the corresponding tube is given by $d=\sqrt{m^2+n^2+mn}\cdot\SI{0.0785}{\nano\metre}$. We study armchair ($m,m$)-CNTs with chiral indices $m=4$ up to $m=10$, which are equivalent to diameters $d=\SI{0.54}{\nano\metre}$ up to $d=\SI{1.4}{\nano\metre}$. We apply the algorithm described in \sref{NJP:Theory} to calculate transmission spectra of defective CNTs. Therefore, we subdivide the CNT into pieces such that the subsystems in \fref{NJP:fig:transport:scheme}b consist of either an ideal CNT unit cell or a defect cell.

We consider three different defect types, which are shown exemplarily for the (4,4)-CNT in \fref{NJP:fig:defect:geometry} in comparison with the defect-free unit cell. The first defect is a monovacancy (MV), where simply one atom was removed. No geometry optimization was performed. The cell of this defect consists of one unit cell. This structure is rather academic, because it has three dangling bonds, which would hardly persist in reality. The second defect is a monovacancy, where the dangling bonds of the adjacent atoms are saturated with hydrogen (MV$_\text{3H}$). The positions of the hydrogen atoms are optimized. This defect consists of three unit cells because of the larger extension which is necessary due to the hydrogen atoms. The third defect type considered is a divacancy (DV), where two adjacent atoms are removed, and the geometry of the whole cell is optimized. The cell of this defect consists of three or four unit cells depending on the lateral position of the removed atoms. Two different orientations are possible: perpendicular (DV$_\text{perp}$) and diagonal (DV$_\text{diag}$). Furthermore, there are $4m$ different positions for the MV, the MV$_\text{3H}$, and the DV$_\text{diag}$, and $2m$ positions for the DV$_\text{perp}$, where $m$ is the chiral index of the tube.

The geometry optimization was performed using DFT, as implemented in Atomistix ToolKit~\cite{ATK.12.8.2, PhysRevB.65.165401}. We used the generalized gradient approximation of Perdew and Zunger~\cite{PhysRevB.23.5048}, norm-conserving Troullier-Martins pseudopotentials~\cite{PhysRevB.43.1993}, and a double zeta plus double polarization (DZDP) basis set of the SIESTA type \cite{JPhysCondMat.14.2745}.

The central region of the CNTs consists of $N$ cells, where we choose $N=\num{1000}$ for the MV and the DV and $N=\num{10000}$ for the MV$_\text{3H}$. To describe CNTs with defects, we choose a constant defect probability $p'_\text{D}$ (per cell), which determines for each of the $N$ cells whether it is a defect cell or not. If it is a defect cell, one of the $4m$ (MV and MV$_\text{3H}$) respectively $6m$ (DV) positions is chosen randomly. We vary the defect probability between $p'_\text{D}=10^{-1}$ and  $p'_\text{D}=10^{-4}$. The absolute number of defects in the device is $N_\text{D}$ and the length of the central region is $L = N_\text{D} L_X + (N-N_\text{D}) L_\text{UC}$, where $L_\text{UC}$ is the length of the ideal unit cell and $L_X$ is the length of the defect cell of type $X\in\left\{\text{MV}, \text{MV}_\text{3H}, \text{DV}\right\}$. As a consequence of the fact that the defect cells can be longer than one unit cell, the length of the central region depends on $p'_\text{D}$. While using $p'_\text{D}$ within the algorithm, a natural definition of the defect probability is based on the total length of the respective central region, i.e. number of defects per equivalent number of ideal unit cells. Thus, we use $p_\text{D}=\frac{N_\text{D}}{L/L_\text{UC}}$ for the evaluation of our results.

For the statistical description we consider an ensemble of configurations. For the limit $T=\SI{0}{\kelvin}$ we choose \num{10000} configurations. For temperatures above $T=\SI{100}{\kelvin}$, \num{1000} configurations are sufficient. By averaging over all configurations with the same $p_\text{D}$ we have an average number of defects\footnote{Note that in the results section where we discuss the length and probability dependence, we average results over configurations with same $p_\text{D}$ and plot them as a function of $\left\langle N_\text{D}\right\rangle$. In the diagrams where we discuss the dependence on the number of defects, we average results over configurations with same $N_\text{D}$ and plot them as a function of $N_\text{D}$. However, in both cases the CNTs are created in the same way, i.e. by choosing a constant defect probability. For the determination of the localization length we use $N_\text{D}$.} $\left\langle N_\text{D}\right\rangle=p'_\text{D}N=p_\text{D}L/L_\text{UC}$. Consequently, the absolute number of defects in the central region is distributed binomially and the defect distance is distributed nearly exponentially. Average transmissions and the average conductances are calculated by means of arithmetic averages. As we do not mix defects, our CNTs contain only one defect type and the obtained results are characteristic for the corresponding type.

\articlesection{Results and discussion}\label{NJP:Results}

Firstly, we will discuss the influence of a single defect within an otherwise perfect CNT. Secondly, the average transport properties of randomly distributed defects will be studied and localization lengths will be derived from the length dependence of the conductance. Thirdly, we will present the dependence of the localization length on the CNT diameter and the temperature.

\begin{figure}[t]
	\includegraphics{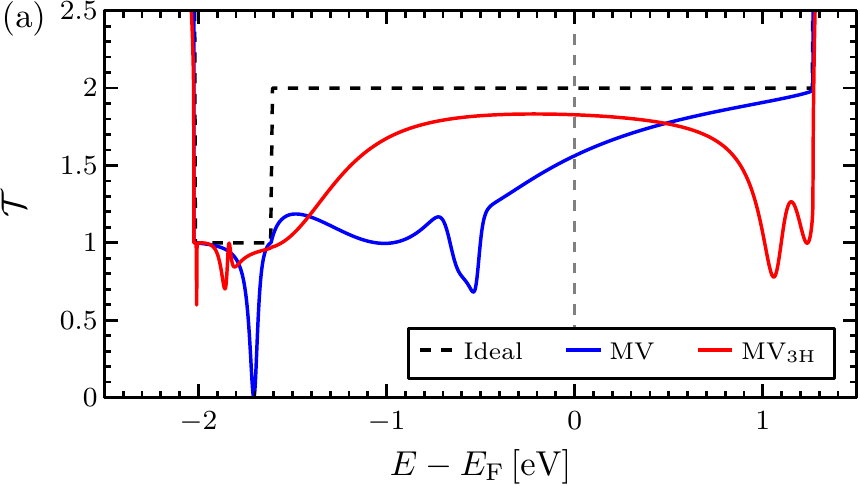}\hfill
	\includegraphics{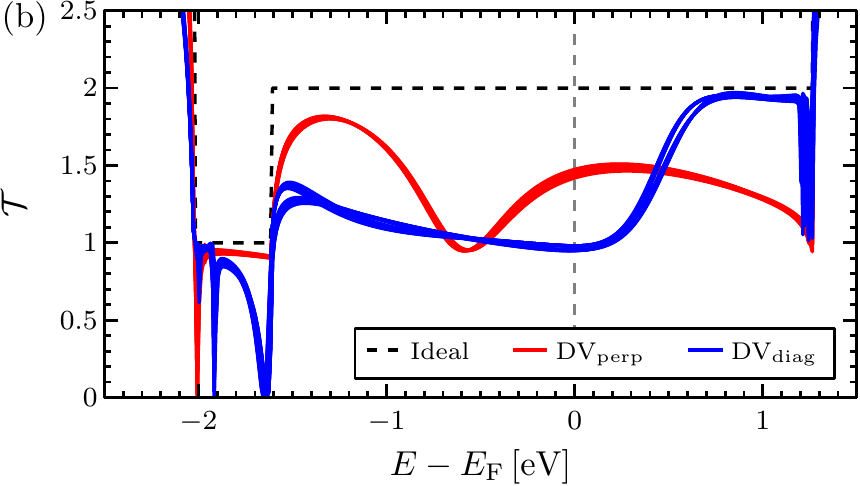}
	\caption[Transmission spectrum of a (4,4)-CNT with a single defect]{Transmission spectrum of a (4,4)-CNT with a single MV and a single $\text{MV}_\text{3H}$ (a) and a single DV in perpendicular and diagonal orientation (b). For each defect several curves of the ensemble are plotted (16 for MV, MV$_\text{3H}$ and DV$_\text{diag}$, 8 for DV$_\text{perp}$). The spectra are nearly identical due to symmetry.}\label{NJP:fig:T:single}
\end{figure}

\articlesubsection{Single defects}\label{NJP:Single}

The defect geometry itself has a non-negligible influence on the electronic properties of CNTs, even in the case of small atomic displacements. So we first compare our results of a single defect with previous work \cite{JPhysCondMat.20.294214}. Here, the central region of the RGF consists of only one defect cell.

The transmission spectra of (4,4)-CNTs with a single MV, MV$_\text{3H}$, and DV are shown in \fref{NJP:fig:T:single} for all possible positions. The transmission of an ideal CNT without defects represents the ballistic limit $G_\text{bal}=2\,\text{G}_0$ (i.e. the transmission at the Fermi level is $\transmission_\text{bal}(E_\text{F})=2$). Due to rotation and mirror symmetry of the system, the curves for the different defect positions are nearly identical, as expected. Whereas the MV at $T=\SI{0}{\kelvin}$ has a conductance of $G_\text{MV}\approx\num{1.6}\,\text{G}_0$, the conductance of the MV$_\text{3H}$ has a larger value of $G_{\text{MV}_\text{3H}}\approx\num{1.8}\,\text{G}_0$. The DV has a smaller conductance than the MV and the MV$_\text{3H}$ due to the larger extension of the DV. Both DV types show different behaviour depending on their orientation. $G_{\text{DV}_\text{diag}}\approx\num{1.0}\,\text{G}_0$ is smaller than $G_{\text{DV}_\text{perp}}\approx\num{1.5}\,\text{G}_0$ due to the larger extension of the DV$_\text{diag}$ along the circumference of the tube.

The results of the DV$_\text{diag}$ and the DV$_\text{perp}$ split into a family of curves corresponding to the different orientations of single defects. This can be explained by the fact that after the geometry optimization the whole cell slightly deviates from perfect symmetry. This leads to non-perfect connections to the neighbour cells. Numerical inaccuracies also play a role. Overall, these deviations are sufficiently small.

The qualitative trend as well as the quantitative values of MV, MV$_\text{3H}$ and DV are in accordance with~\cite{JPhysCondMat.20.294214}. It is particularly noteworthy for the MV$_\text{3H}$, given the fact that the treatment of the dangling bonds is different (no passivation but optimization in~\cite{JPhysCondMat.20.294214}, no optimization but passivation here).

It can be summarized, that a single mono- or divacancy defect reduces the conductance of a (4,4)-CNT by about 10\% to 50\%.

\articlesubsection{Randomly distributed defects}\label{NJP:Random}

\begin{figure}[t]
	\includegraphics{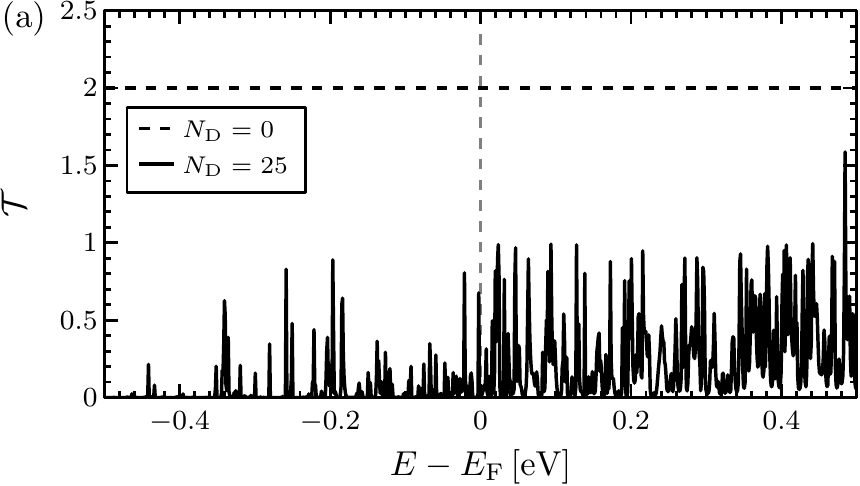}\hfill
	\includegraphics{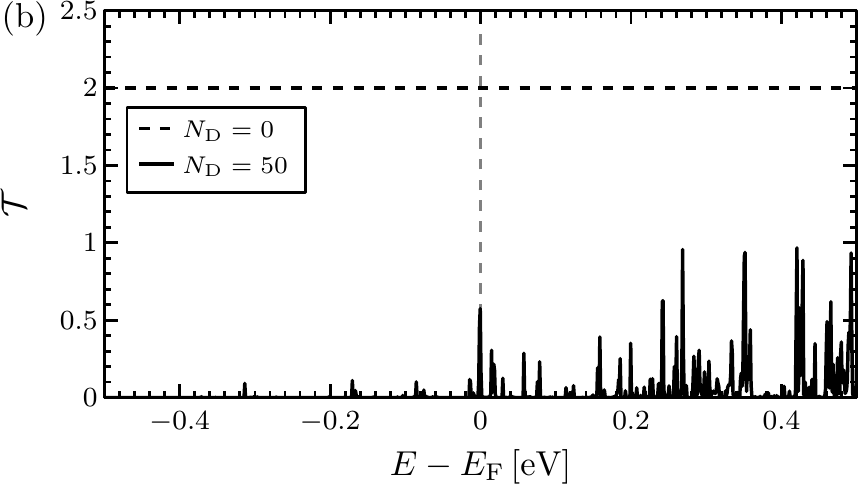}
	\caption[Transmission spectrum of a (4,4)-CNT with random defects]{Transmission spectrum of an individual (4,4)-CNT with 25 (a) and 50 (b) randomly distributed MV defects within \num{1000} cells. See supplementary data for $N_\text{D}=\num{10}$ and $N_\text{D}=\num{100}$.}\label{NJP:fig:T:random}
\end{figure}

\Fref{NJP:fig:T:random} shows the transmission spectra of a (4,4)-CNT with 25 and 50 randomly distributed MV defects within \num{1000} cells. The curves can be explained as follows: If the CNT contains more than one defect, the transmission is not just lowered, it is also characterized by multiple sharp peaks which occur due to resonances caused by constructive interference of scattered electrons. This effect gets stronger with increasing number of defects. If we consider a fixed CNT length with a fixed number of defects the effect gets also stronger with larger distances between the defects. This can be easily understood in the simple one-dimensional picture of an electron wave, scattered by multiple potential barriers. In the case of two barriers \cite{arXiv.1210.0970} it can be compared to a Fabry-P\'erot interferometer. Since the resonances depend strongly on the actual configuration -- that means on the type and structure of the defects and their distances and relative orientations -- the resonances are randomly distributed \cite{PhysRevB.27.831}.

\begin{figure}[b!]
	\centering
	\begin{minipage}{\OneColumnWidth}(a) MV\end{minipage}\hfill
	\begin{minipage}{\OneColumnWidth}(b) MV$_\text{3H}$\end{minipage}\\[0.5em]
	\includegraphics{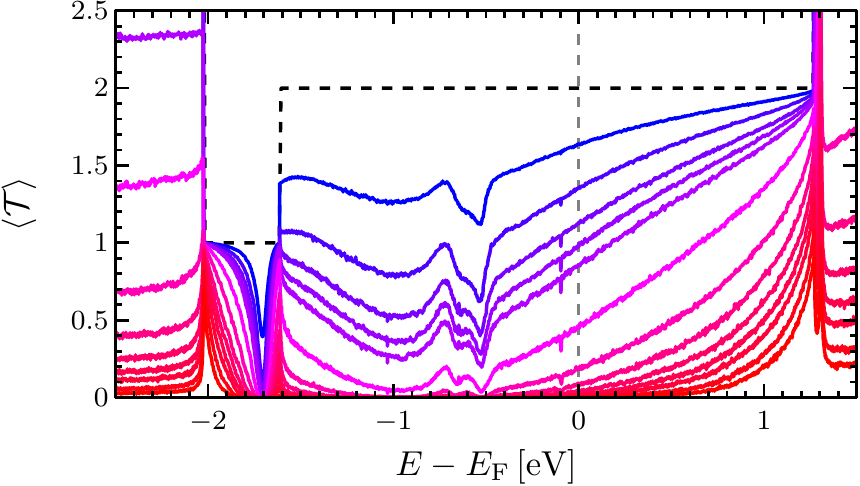}\hfill
	\includegraphics{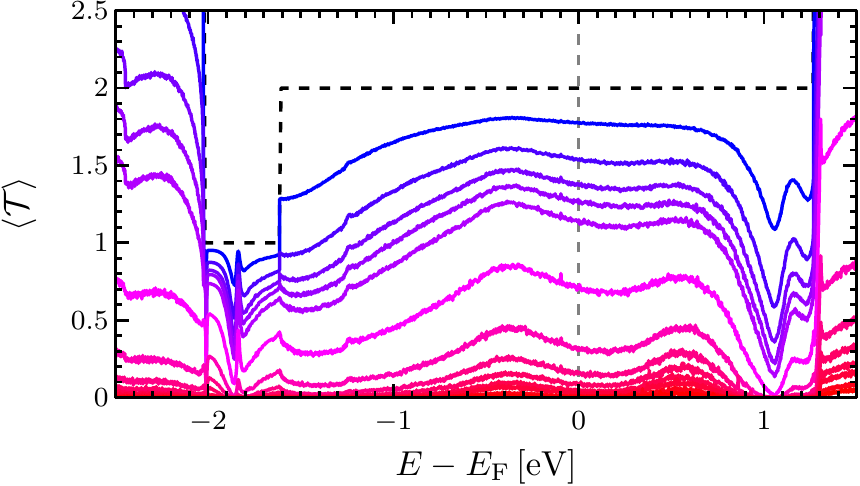}\\[1em]
	\begin{minipage}{0.6\textwidth}
		(c) DV\\[0.5em]
		\includegraphics{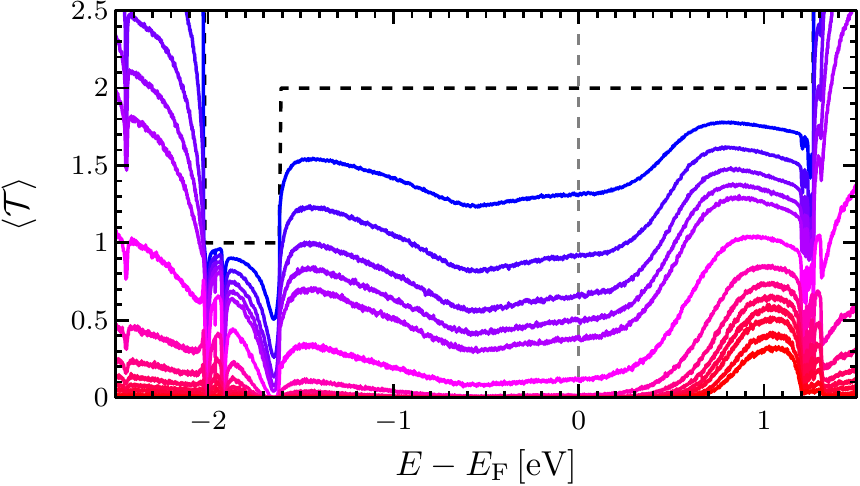}
		\includegraphics{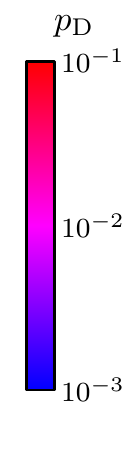}
	\end{minipage}
	\caption[Average transmission spectra of (4,4)-CNTs with random defects]{Average transmission spectra of randomly distributed MV (a), $\text{MV}_\text{3H}$ (b), and DV defects (c) in a (4,4)-CNT with different defect probabilities $p_\text{D}$. The length of the CNT is \num{1000} cells. The average was performed over \num{1000} configurations.}\label{NJP:fig:T:avg}
\end{figure}

By looking at a series of such pictures with increasing numbers of defects (figures \ref{NJP:fig:T:random}a and \ref{NJP:fig:T:random}b) we see that the maximum height of the peaks is only slightly reduced. Large peaks remain at $G\approx 1\,\text{G}_0$ (besides numerical artifacts due to the finite energy resolution), but the resonances are getting sharper, which leads to a smaller number of peaks in the calculated transmission spectrum due to the finite energy resolution. For small temperatures the spiky structure of the transmission spectrum leads to a spreading of the conductance values over many orders of magnitude depending on the individual configurations. The conductance of an individual configuration is unexpectedly high in case of a peak at the Fermi level. For example the configuration of \fref{NJP:fig:T:random}a (25 defects) has $\transmission=\num{0.013}$ whereas the configuration of \fref{NJP:fig:T:random}b (50 defects) has a higher value of $\transmission=\num{0.37}$ although it contains more defects. To describe this behaviour we look at an ensemble of different defect realizations.

\Fref{NJP:fig:T:avg} shows the ensemble average (\num{1000} defect configurations) of the transmission spectra for the three different defect types. The individual resonances vanish and, irrespective of the defect probability, the transmission spectra show a similar behaviour and they are strongly decreasing with increasing defect probability. Their features are surprisingly similar to the case of one defect.

\begin{figure}[t]
	\begin{minipage}{\OneColumnWidth}(a) $T=\SI{0}{\kelvin}$\end{minipage}\hfill
	\begin{minipage}{\OneColumnWidth}(b) $N_\text{D}=\num{100}$\end{minipage}\\[0.5em]
	\includegraphics{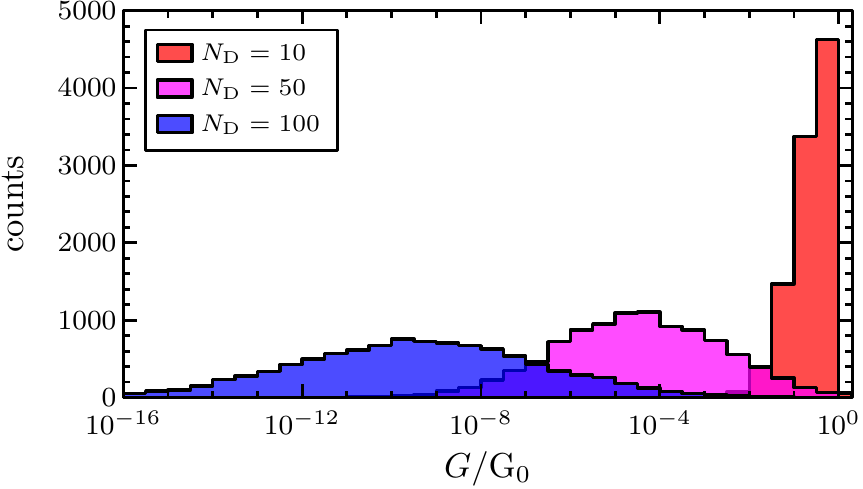}\hfill
	\includegraphics{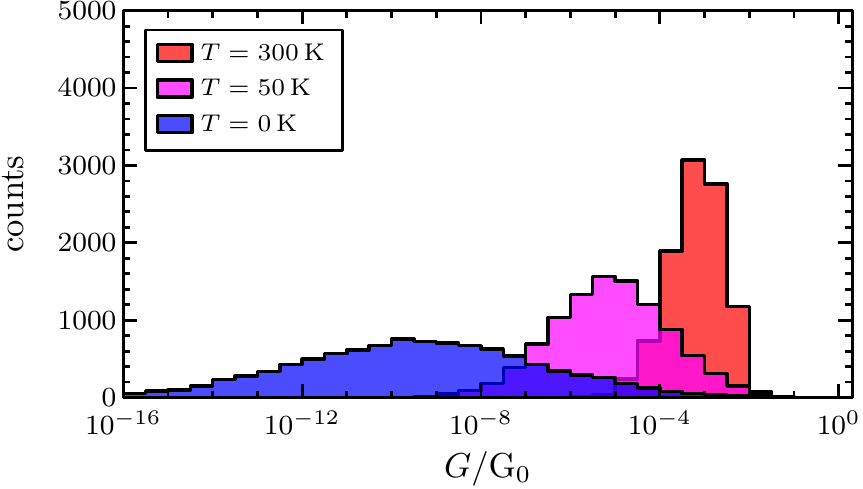}
	\caption[Conductance histogram for (10,10)-CNTs with random defects]{Conductance histogram for an ensemble of \num{10000} configurations of (10,10)-CNTs with different numbers of DV defects at $T=\SI{0}{\kelvin}$ (a) and different temperatures at $N_\text{D}=100$ (b). The defect probability is $p_\text{D}\approx\num{0.075}$.}\label{NJP:fig:G:distribution}
\end{figure}

\Fref{NJP:fig:G:distribution}a shows the conductance histogram for an ensemble of \num{10000} configurations of (10,10)-CNTs with different numbers of DV defects at $T=\SI{0}{\kelvin}$. The conductance values are log-normally distributed in the limit of large disorder (for a detailed description of such distributions, their moments and cumulants, see \cite{PhysRevB.72.195407}). The small disorder case differs from that because of the ballistic limit $G_\text{bal}=2\,\text{G}_0$. Furthermore, the spread of the conductance values increases strongly with increasing number of defects due to the previously mentioned fact that the transmission peaks become sharper. In that case, a large ensemble is necessary to get a reliable average.

\Fref{NJP:fig:G:distribution}b shows the same conductance histogram for different temperatures and $N_\text{D}=\num{100}$. It follows from \eref{NJP:eqn:G} that the conductance of an individual configuration will grow strongly with increasing temperature as long as the transmission spectrum has no peak at the Fermi level, which is so in most cases. Consequently, the distribution tail of small conductances and the average conductance are shifted to higher conductances and the width of the distribution decreases with increasing temperature.

\articlesubsection{Localization exponent}\label{NJP:L}

\begin{figure}[b!]
	\centering
	\includegraphics{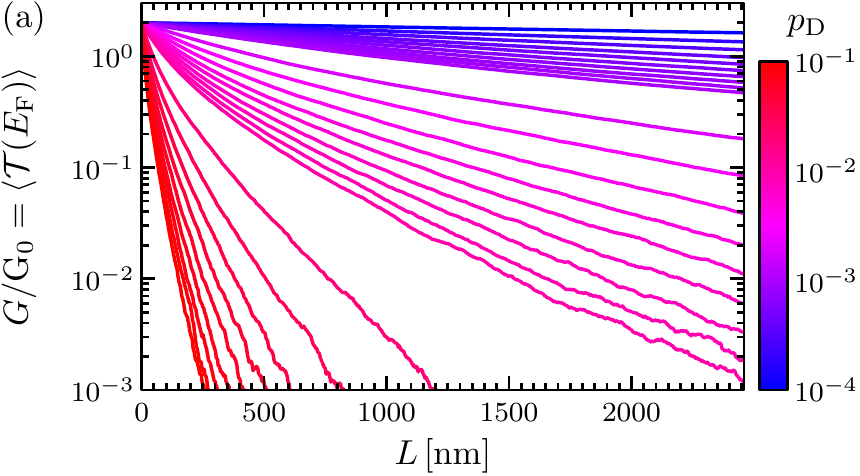}\hfill
	\includegraphics{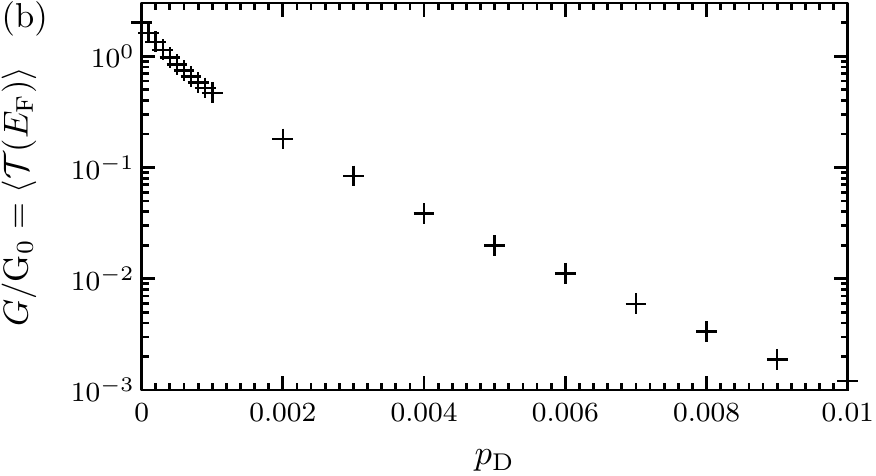}\\[1em]
	\includegraphics{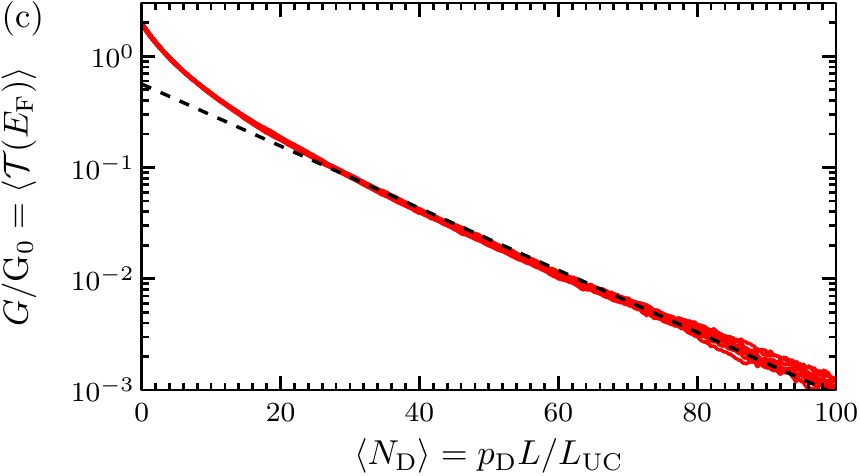}
	\caption[Conductance as a function of length, defect probability and number of defects]{Conductance of MV defects in a (4,4)-CNT. Dependence on the CNT length for different defect probabilities (a), dependence on the defect probability for \num{10000} cells (b), and dependence on the number of defects (c, solid red curve). The dashed black curve is an exponential regression for the interval $\avg{N_\text{D}}=\num{20}\ldots\num{100}$.}\label{NJP:fig:G:L,p,N}
\end{figure}

To determine the scaling behaviour of the previously mentioned reduction of the transmission for increasing defect probabilities, we calculate the conductance \eref{NJP:eqn:G:0K} at $T=\SI{0}{\kelvin}$ in dependence on the length and defect probability of the defective region of the CNT. We average over \num{10000} configurations with up to \num{10000} cells in the central region. The result is shown in \fref{NJP:fig:G:L,p,N}a for a (4,4)-CNT with MV defects. Here, the length is given in terms of the number of cells $N$ in the central region. For long CNTs an exponential decrease of the conductance is observed. This finding also holds for the other CNTs and defect types.

The decrease of the conductance can be interpreted using the Anderson model of localization \cite{PhysRev.109.1492, Mello2004, PhysRevLett.47.1546, PhysRevLett.42.673}, which states that the conductance for fixed disorder strength scales exponentially with the tube length $L$, which is proportional to the number of cells $N$,
\begin{articleequation}
	G \simeq \text{e}^{-L/\ell_\text{loc}} \qquad , \label{NJP:eqn:loc:L}
\end{articleequation}%
where the localization length $\ell_\text{loc}$ is a characteristic parameter of the system.

Plotting the dependence of the conductance on the defect probability as shown in \fref{NJP:fig:G:L,p,N}b, a similar behaviour is observed. Hence, it is an obvious assumption that all curves of \fref{NJP:fig:G:L,p,N}a could fall together on an universal curve by scaling them with $p_\text{D}$. This assumption is clearly confirmed by \fref{NJP:fig:G:L,p,N}c, which concentrates all the data of figures \ref{NJP:fig:G:L,p,N}a and \ref{NJP:fig:G:L,p,N}b into one single curve. Thus, the number of defects is the relevant scaling parameter of the conductance.

Corresponding to the Anderson model, we have
\begin{articleequation}
	G \simeq \text{e}^{-N_\text{D}/N_\text{D}^\text{loc}} \label{NJP:eqn:loc:N}
\end{articleequation}%
in the limit of a large number of defects and a characteristic localization exponent $N_\text{D}^\text{loc}$ can be extracted from the data (black line in \fref{NJP:fig:G:L,p,N}c). It follows from \eref{NJP:eqn:loc:L} and \eref{NJP:eqn:loc:N} that $N_\text{D}^\text{loc}\propto\ell_\text{loc}p_\text{D}$, which means that the localization length is inversely proportional to the defect probability. For the depicted example (MV defects) we get $N_\text{D}^\text{loc}=17$. The MV$_\text{3H}$ has $N_\text{D}^\text{loc}=23$ and the DV has $N_\text{D}^\text{loc}=6.0$. If we consider a moderate defect probability of $p_\text{D}=\num{0.01}$ (i.e. one defect every $\SI{25}{\nano\metre}$), we get localization lengths $\ell_\text{loc}=\SI{420}{\nano\metre}$ for the MV, $\ell_\text{loc}=\SI{570}{\nano\metre}$ for the MV$_\text{3H}$, and $\ell_\text{loc}=\SI{150}{\nano\metre}$ for the DV. This agrees with typical localization lengths of a few hundred nanometres \cite{NatureMaterials.4.534}.

\articlesubsection{Diameter dependence and temperature dependence of the localization exponent}\label{NJP:D}

\begin{figure}[t]
	\includegraphics{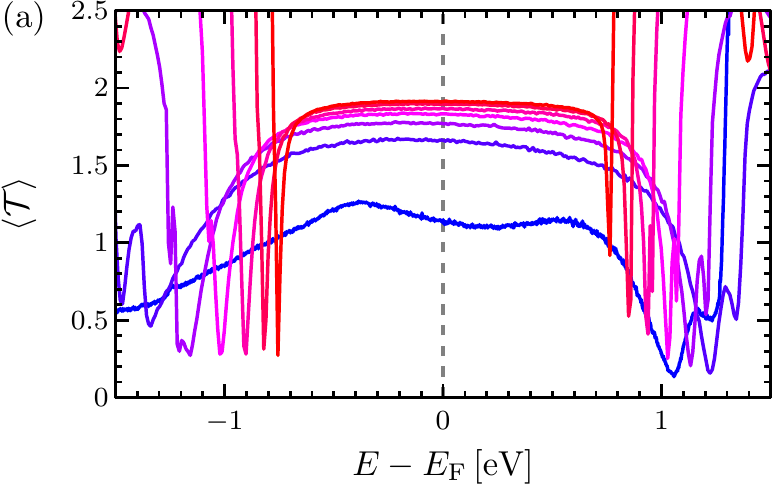}\hfill
	\includegraphics{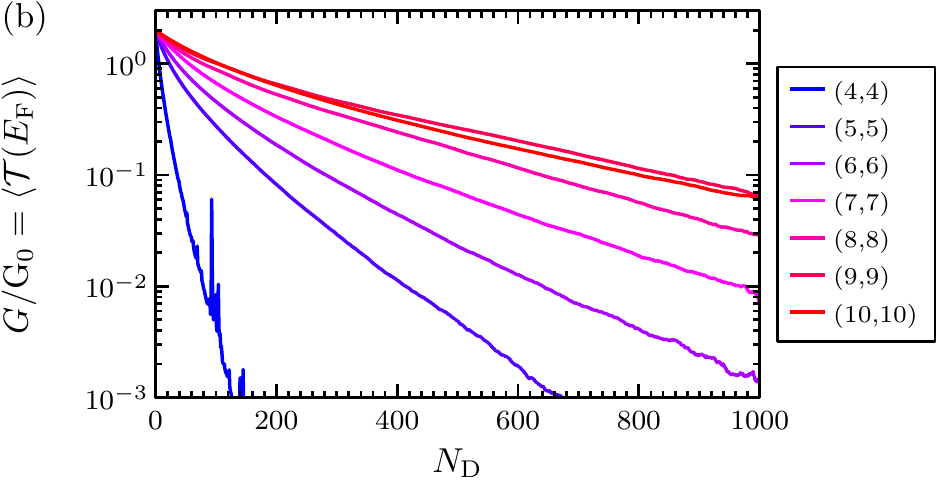}
	\caption[Transmission spectra and conductance as a function of the number of defects]{Average transmission spectra (\num{1000} configurations, \num{1000} cells, $p_\text{D}\approx\num{0.005}$) of randomly distributed MV$_\text{3H}$ defects (a) and conductance as a function of the number of MV$_\text{3H}$ defects (b) for different CNTs. See supplementary data for MV and DV defects.}\label{NJP:fig:d:MV3H}
\end{figure}

So far, most of the data were shown for the (4,4)-CNT, which has a diameter $d=\SI{0.54}{\nano\metre}$. To get a more comprehensive view we have performed the calculations systematically for bigger armchair-CNTs up to the (10,10)-CNT, which has a diameter $d=\SI{1.4}{\nano\metre}$. \Fref{NJP:fig:d:MV3H}a depicts average transmission spectra of armchair CNTs with increasing diameter containing MV$_\text{3H}$ defects. It shows a strong increase of the transmission due to the fact that the number of atoms per unit cell increases linearly with the CNT diameter so that the relative size of a defect is lowered. For the (9,9)- and the (10,10)-CNT the transmission around the Fermi energy approaches the theoretical maximum of $\transmission_\text{bal}=2$.

As in \fref{NJP:fig:G:L,p,N}c, \fref{NJP:fig:d:MV3H}b shows an exponential decrease of the conductance with the number of defects (in the limit of many defects) for all CNTs. We observe the same behaviour also for the DV, but there the conductance is much smaller compared to the case of MV$_\text{3H}$ defects, because the DV has a much bigger influence.

\begin{figure}[t]
	\includegraphics{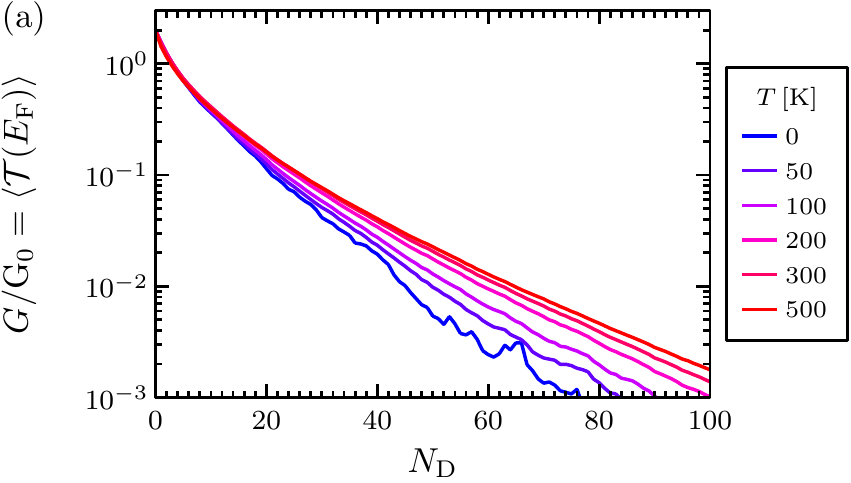}\hfill
	\includegraphics{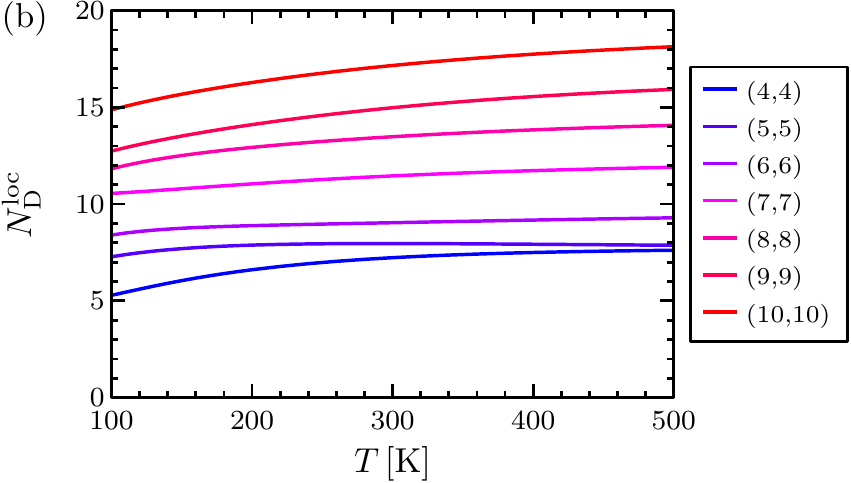}
	\caption[Conductance and localization exponent as a function of temperature]{Conductance (average over \num{1000} configurations, $p_\text{D}\approx\num{0.081}$) of a (10,10)-CNT in dependence on the number of DV defects at different temperatures (a) and localization exponent in dependence on the temperature for different CNTs with DV defects (b). See supplementary data for MV and MV$_\text{3H}$ defects.}\label{NJP:fig:Temperature}
\end{figure}

\begin{figure}[t]
	\centering
	\includegraphics{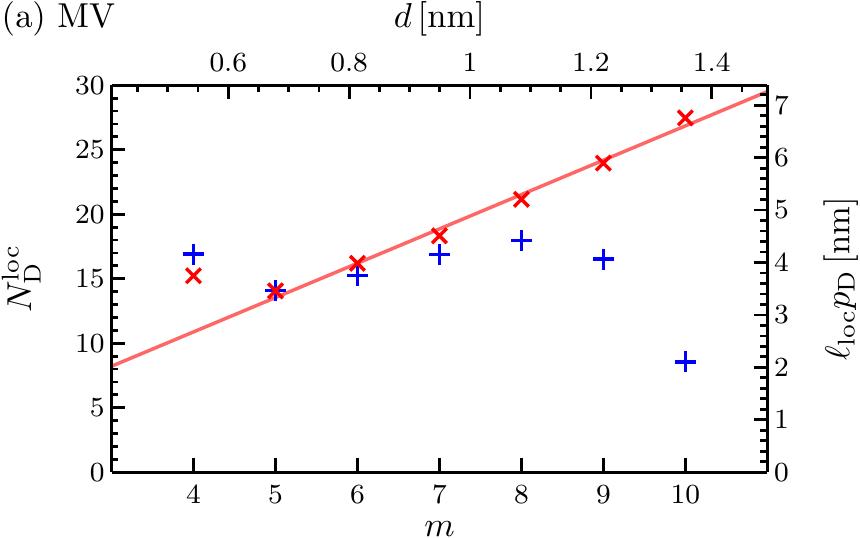}\hfill
	\includegraphics{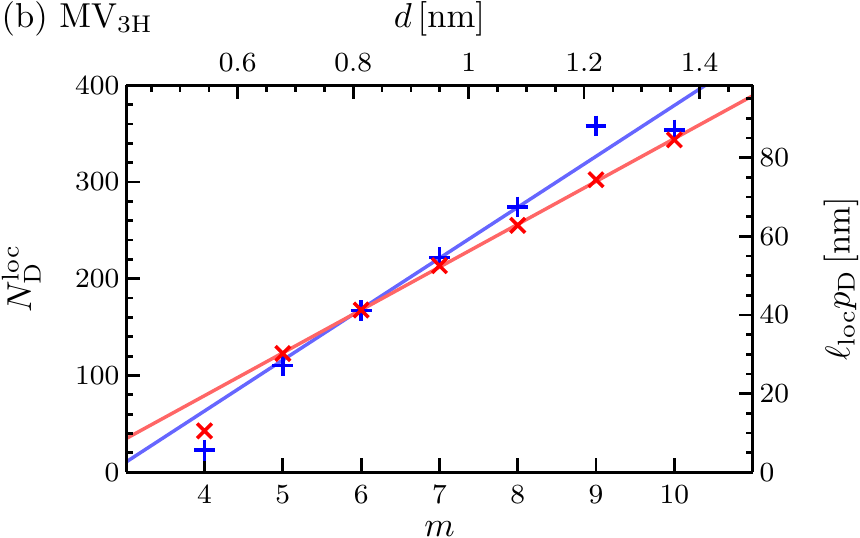}\\[1em]
	\includegraphics{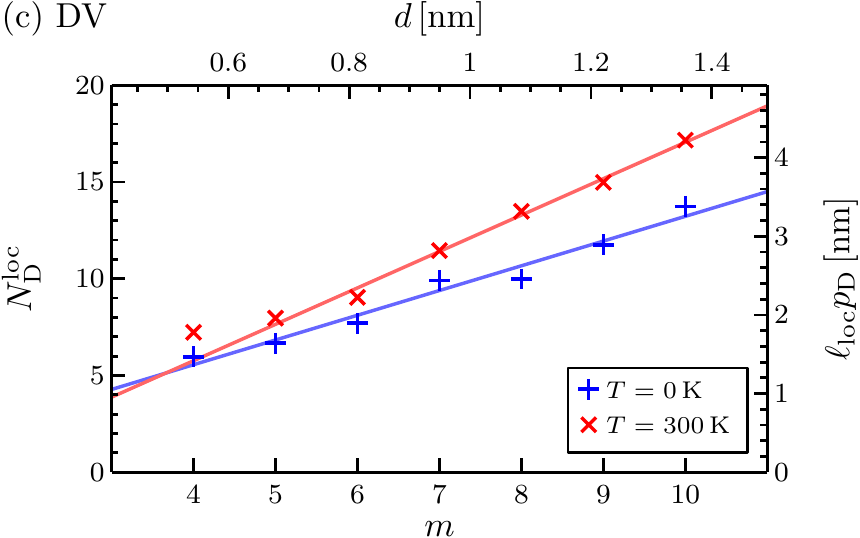}
	\caption[Diameter dependence of the localization length]{Diameter dependence of the localization length for CNTs with MV (a), MV$_\text{3H}$ (b), and DV defects (c). The results are shown at $T=\SI{0}{\kelvin}$ (blue, $+$) and at $T=\SI{300}{\kelvin}$ (red, $\times$). The solid lines are linear regressions in the range $m=5\ldots 10$ (see text for details).}\label{NJP:fig:lloc}
\end{figure}

So far, the conductance at $T=\SI{0}{\kelvin}$ has been studied. According \eref{NJP:eqn:G}, the influence of finite temperature is included by convolving the transmission spectrum with the Fermi distribution. The result is shown in \fref{NJP:fig:Temperature}a, where the conductance of a (10,10)-CNT with DV defects is drawn as a function of the number of defects for different temperatures. Hence, the influence of the temperature on the localization exponent can directly be seen in this picture. The different curves show that the conductance increases with temperature. This happens in such a way that the localization length increases, too. This can be explained by the fact that, in general, the scaling behaviour of the transmission spectrum is energy dependent. So the localization exponent defined via \eref{NJP:eqn:loc:N} is temperature dependent.

We proceed by studying the temperature dependence of the localization exponent for different armchair CNTs with DV defects explicitly (see \fref{NJP:fig:Temperature}b). The localization exponent increases for all CNTs with temperature, but the dependence is rather weak. The influence of the tube diameter, denoted by different curves, is much larger than the effect of temperature.

The same behaviour as in figures \ref{NJP:fig:Temperature}a and \ref{NJP:fig:Temperature}b can be seen for the MV defect. In contrast, all the $G(N_\text{D})$ curves for different temperatures fall together for the MV$_\text{3H}$ defect (see figure 4b in supplementary data). This is due to the fact that the corresponding transmission spectrum is nearly constant in a wide range around the Fermi energy (see also figures \ref{NJP:fig:T:single}a and \ref{NJP:fig:d:MV3H}a). It follows that there is no $N_\text{D}^\text{loc}(T)$ dependence for the MV$_\text{3H}$ defect.

Analyzing the data shown in \fref{NJP:fig:Temperature}b at fixed temperature, we can conclude that CNTs with different diameters have different localization exponents. The figure already indicates a rather linear scaling of $N_\text{D}^\text{loc}$ with the chiral index of the tube. This dependence is discussed in detail in the following.

\Fref{NJP:fig:lloc} shows the localization exponent as a function of the chiral index $m$, which is proportional to the tube diameter $d$. The three defect types and two different temperatures are considered for comparison. At $T=\SI{0}{\kelvin}$ a linear dependence can be seen for the two defect types MV$_\text{3H}$ and DV, which was also found in \cite{JPhysCondMat.20.304211}. Not only the conductance itself is much smaller for the DV than for the MV$_\text{3H}$, but also the localization parameter is smaller by a factor of about 20. This means that one additional DV defect has the same effect as 20 additional MV$_\text{3H}$ defects, which is quite surprising. The localization exponents of the MV defect show no clear trend because of the deep valley in the transmission spectrum which shifts through the Fermi energy with increasing CNT diameter (see figure 3a in supplementary data).

The localization exponents calculated for $T=\SI{300}{\kelvin}$ are also displayed in \fref{NJP:fig:lloc}. Again we see the linear dependence for the DV but with a larger slope, leading to the previously mentioned higher localization exponent. The MV$_\text{3H}$ shows nearly the same results as for $T=\SI{0}{\kelvin}$ because of the fact that the transmission spectrum is nearly constant around $E_\text{F}$ (see \fref{NJP:fig:d:MV3H}a) and thus, the temperature dependence is rather small. Furthermore, a linear behaviour can now be seen for the MV, too. This is due to the fact that the broader convolution kernel (see \eref{NJP:eqn:G}) at higher temperature reduces the influence of the very narrow but deep valley in the transmission spectrum at the Fermi energy.

Though we get the linear dependence $N_\text{D}^\text{loc}(d)$ for large diameters, the (4,4)-CNT differs from that finding due to the fact, that it has a rather small diameter and thus curvature effects dominate the electronic structure. The  different shapes of the transmission spectra in \fref{NJP:fig:d:MV3H}a already confirm this. Consequently, $m=4$ will be omitted in the quantification of $N_\text{D}^\text{loc}(d)$. Furthermore, all detailed features in the transmission spectrum cause changes in the localization exponent. The MV (\fref{NJP:fig:lloc}a) demonstrates this.

A regression of the previously discussed linear dependence at $T=\SI{0}{\kelvin}$ yields
\begin{articleequation}
	\begin{aligned}
	\ell_\text{loc} &= (\SI{-36}{nm}+\num{95}d)/p_\text{D} &&\quad\text{for the MV}_\text{3H}~\text{and} \notag\\
	\ell_\text{loc} &= (\SI{0.11}{nm}+\num{2.3}d)/p_\text{D} &&\quad\text{for the DV}.\notag
	\end{aligned}
\end{articleequation}
The diameter dependence at $T=\SI{300}{\kelvin}$ is characterized by
\begin{articleequation}
	\begin{aligned}
	\ell_\text{loc} &= (\SI{0.1}{nm}+\num{4.8}d)/p_\text{D} &&\quad\text{for the MV}, \notag\\
	\ell_\text{loc} &= (\SI{-24}{nm}+\num{80}d)/p_\text{D} &&\quad\text{for the MV}_\text{3H},~\text{and} \notag\\
	\ell_\text{loc} &= (\SI{-0.4}{nm}+\num{3.4}d)/p_\text{D} &&\quad\text{for the DV}.\notag
	\end{aligned}
\end{articleequation}
In conclusion we derived an expression which allows us to predict the localization lengths and thus the scaling behaviour of the conductance of armchair-CNTs with arbitrary diameter.

\begin{figure}[b!]
	\centering
	\includegraphics{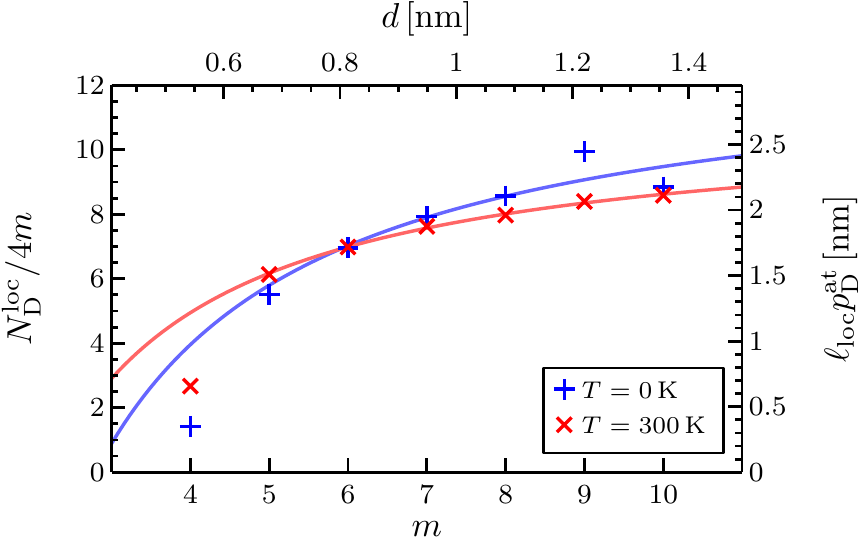}
	\caption[Localization length for constant defect probability per atom]{Diameter dependence of the localization length for constant defect probability per atom $p_\text{D}^\text{at}$ and CNTs with MV$_\text{3H}$ defects. The data points ($+$ and $\times$) and the regressions (solid lines) are the same as in \fref{NJP:fig:lloc}b. See supplementary data for MV and DV defects.\\~}\label{NJP:fig:lloc2}
\end{figure}

At the same time we have to keep in mind that, while keeping $p_\text{D}$ constant, an increase of the diameter leads to an increasing number of atoms per unit cell. So the limit $d\rightarrow\infty$ leads to $\ell_\text{loc}\rightarrow\infty$ and consequently $G\rightarrow 2\,\text{G}_0$. This is clear because the structure tends to a CNT with infinite diameter, where a single defect has no influence. On the other hand, a constant defect density implies a constant defect probability per atom $p_\text{D}^\text{at}=p_\text{D}/4m$. To address this, figure~\ref{NJP:fig:lloc2} shows $\ell_\text{loc}p_\text{D}^\text{at}$ as a function of the tube diameter. It can be seen that the data points approach a constant value. For tubes with large diameter this leads to a universal and diameter independent localization length, which is simply the slope of the linear regression in \fref{NJP:fig:lloc}. In such a scenario, the conductance is independent of the diameter of the CNT, and we approach the case of defective graphene.

As discussed before, the localization exponents of the three studied defect types differ widely. Especially the one of the MV$_\text{3H}$ defect is higher by a factor of 20 than the one of the DV defect. The reason for this is explained in the following. An obvious assumption is that the localization exponent depends systematically on a parameter describing the specific influence of the defect. Concerning Anderson disorder with an uniform disorder distribution, the localization length is related to the disorder strength $W$ as $1/W^2$ \cite{Thouless1979}\footnote{This is the relevant part of the Thouless relation. It is valid in the limit of large systems, weak disorder, and away from the band center and the band edges.}. In contrast, the investigation of realistic defects is based on a statistical distribution of the defect positions instead of a disorder strength. Consequently, another measure has to be found. The defect probability is a bad one, because it neglects the defect type. A better measure would be the conductivity of a single defect because it reflects the specific perturbation of the otherwise ballistic electron transport.

\begin{figure}[tb]
	\centering
	\includegraphics{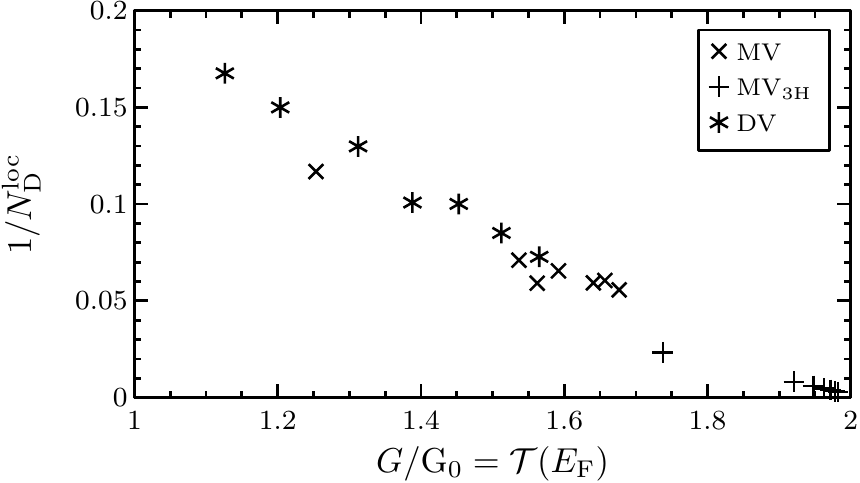}
	\caption[Inverse localization exponent as a function of the single defect conductance]{Inverse localization exponent (data points of \fref{NJP:fig:lloc}) plotted against the conductivity of one single defect (section~\ref{NJP:Single}) at $T=\SI{0}{\kelvin}$. Different symbols correspond to the different defect types MV, MV$_\text{3H}$ and DV. For each defect type, the data points of the (4,4)-CNT up to the (10,10)-CNT are shown.}\label{NJP:fig:llocG}
\end{figure}

To verify this, \fref{NJP:fig:llocG} shows the inverse localization exponent plotted against the conductivity of one single defect at zero temperature for all the previously shown defect types and CNT diameters. A clear linear dependence can be seen. Thus, we can conclude that the localization exponent scales inversely with the conductance of the CNT with one defect. The case of finite temperatures shows the same behaviour. Furthermore, this relation not only holds for one defect type but is consistent for all three defect types. This strongly indicates that conductances of single defects can be used to characterize localization exponents, independent of the defect structures themselves. Thus, the explanation of the big difference between localization exponents of different defect types is reduced to the explanation of the difference of the conductance of single defects. As discussed in \sref{NJP:Single}, one DV defect drastically reduces the conductance whereas the conductance reduction of one MV$_\text{3H}$ defect is much smaller, especially for larger tubes like the (10,10)-CNT (also compare \fref{NJP:fig:d:MV3H}a and figure 3b of supplementary data).

\articlesection{Summary and conclusions}\label{NJP:conclusions}

In summary, we investigated the influence of static disorder on the transport properties of armchair carbon nanotubes. This was done in a systematical way for different tube diameters, defect types, and temperatures with a consistent theory and provides more comprehensive information than former studies \cite{NatureMaterials.4.534, PhysRevLett.95.266801, JPhysCondMat.20.294214, JPhysCondMat.20.304211}.

The disorder was implemented by random distributions of realistic defects. We focused on mono- and divacancies, which are the most common defects arising in technological processes. For the underlying electronic structure we chose a DFTB model, which combines the accuracy of DFT with the simplicity of TB. Within the standard Landauer transport formalism we used the RGF to treat very large systems in an efficient recursive way, which offers linear scaling of the calculation time with the system size. We calculated the average conductance of ensembles of CNTs with fixed defect probabilities and studied the dependence on the number of defects, the tube diameter and the temperature.

We showed explicitly that the relevant scaling parameter is the absolute number of defects $N_\text{D}$, that means the conductance -- in the limit of large disorder -- scales exponentially with $N_\text{D}$ and consequently also exponentially with the tube length and the defect probability. This behaviour confirms that the system is in the regime of strong localization. Localization exponents $N_\text{D}^\text{loc}$ were calculated for different defect types and for different CNTs at $T=\SI{0}{\kelvin}$ as well as at finite temperatures. As a main result our data show a universal linear dependence $N_\text{D}^\text{loc}(d)$ for high temperatures. This leads to a defect-dependent universal parameter describing the localization in arbitrary armchair-CNTs. Furthermore, we showed that the localization exponent is related to the conductance of the single defect in a consistent way for all three studied defect types.

This work helps to understand electronic transport of mesoscopic systems. On the basis of this systematical study, the localization exponent and thus the scaling behaviour of the conductance of even larger CNTs than the ones explicitly considered here, can be predicted. The present theory could be used to gain more (at least qualitative) information about types and distributions of defects in CNTs from experimentally measurable electron transport characteristics.

In future, further dependencies are of interest. The effect of mixed defect types is not considered yet. It has to be tested, if there exists a simple relation, e.g. an average, between the localization length of CNTs with a mixture of defects and the localization lengths of CNTs with a single defect type. The influence of the chirality, different functionalization types or bent CNTs can also be investigated. The verification of the correlation between the localization exponent and the conductance of a single defect for other systems would be of great interest because it gives the opportunity of estimating or even predicting localization exponents only by doing very cheap calculations of the conductance of single defects.

\newpage
\articlesection*{Supplementary material}
\oldaddcontentsline{toc}{section}{Supplementary material}

\newlength{\ColumnCaption}
\setlength{\ColumnCaption}{0.08\textwidth}
\newlength{\ColumnFigure}
\setlength{\ColumnFigure}{0.15\textwidth}
\newlength{\FigureOneCell}
\setlength{\FigureOneCell}{0.3684\ColumnFigure} 
\newlength{\FigureThreeCells}
\setlength{\FigureThreeCells}{0.7895\ColumnFigure} 
\newlength{\FigureFourCells}
\setlength{\FigureFourCells}{\ColumnFigure}
\begin{figure}[h!]
	\begin{minipage}{\ColumnCaption}
		~
	\end{minipage}\hfill
	\begin{minipage}{\ColumnFigure}
		\centering
		(5,5)
	\end{minipage}\hfill
	\begin{minipage}{\ColumnFigure}
		\centering
		(6,6)
	\end{minipage}\hfill
	\begin{minipage}{\ColumnFigure}
		\centering
		(7,7)
	\end{minipage}\hfill
	\begin{minipage}{\ColumnFigure}
		\centering
		(8,8)
	\end{minipage}\hfill
	\begin{minipage}{\ColumnFigure}
		\centering
		(9,9)
	\end{minipage}\hfill
	\begin{minipage}{\ColumnFigure}
		\centering
		(10,10)
	\end{minipage}\\
	\begin{minipage}{\ColumnCaption}
		MV
	\end{minipage}\hfill
	\begin{minipage}{\ColumnFigure}
		\centering
		\includegraphics[width=\FigureOneCell]{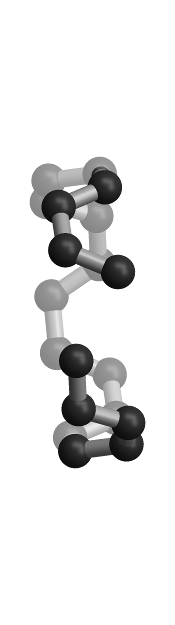}
	\end{minipage}\hfill
	\begin{minipage}{\ColumnFigure}
		\centering
		\includegraphics[width=\FigureOneCell]{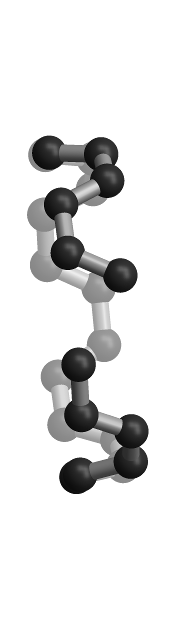}
	\end{minipage}\hfill
	\begin{minipage}{\ColumnFigure}
		\centering
		\includegraphics[width=\FigureOneCell]{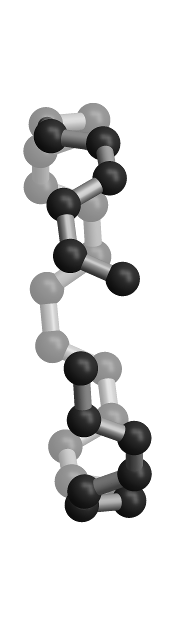}
	\end{minipage}\hfill
	\begin{minipage}{\ColumnFigure}
		\centering
		\includegraphics[width=\FigureOneCell]{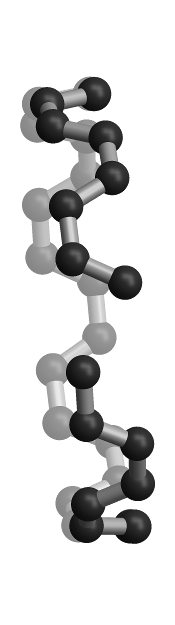}
	\end{minipage}\hfill
	\begin{minipage}{\ColumnFigure}
		\centering
		\includegraphics[width=\FigureOneCell]{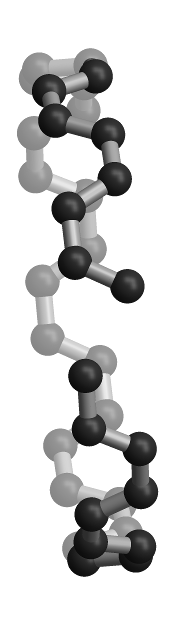}
	\end{minipage}\hfill
	\begin{minipage}{\ColumnFigure}
		\centering
		\includegraphics[width=\FigureOneCell]{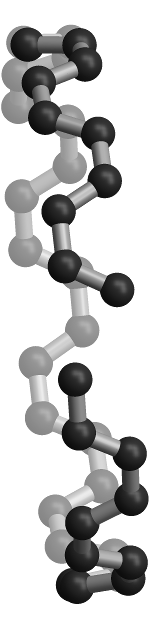}
	\end{minipage}\\
	\begin{minipage}{\ColumnCaption}
		MV$_\text{3H}$
	\end{minipage}\hfill
	\begin{minipage}{\ColumnFigure}
		\centering
		\includegraphics[width=\FigureThreeCells]{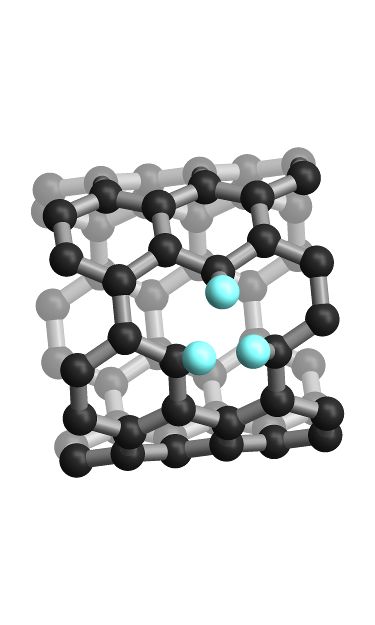}
	\end{minipage}\hfill
	\begin{minipage}{\ColumnFigure}
		\centering
		\includegraphics[width=\FigureThreeCells]{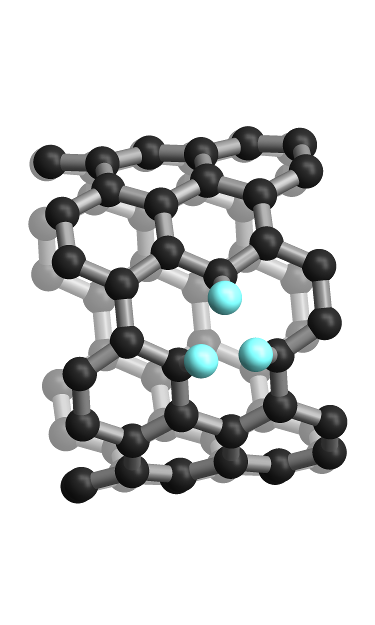}
	\end{minipage}\hfill
	\begin{minipage}{\ColumnFigure}
		\centering
		\includegraphics[width=\FigureThreeCells]{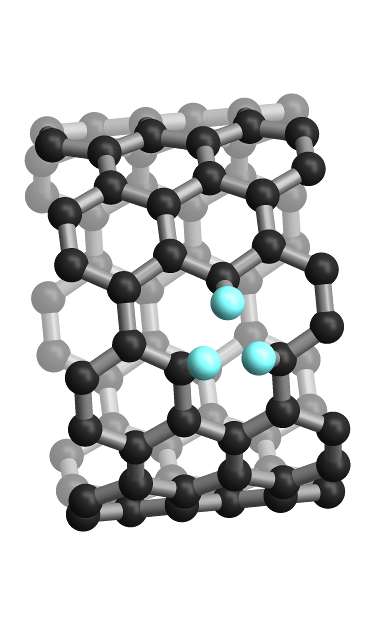}
	\end{minipage}\hfill
	\begin{minipage}{\ColumnFigure}
		\centering
		\includegraphics[width=\FigureThreeCells]{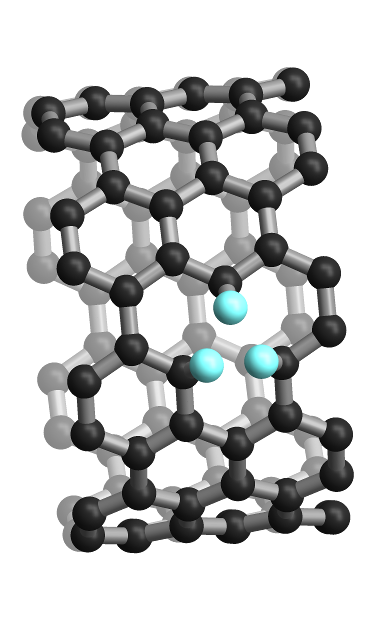}
	\end{minipage}\hfill
	\begin{minipage}{\ColumnFigure}
		\centering
		\includegraphics[width=\FigureThreeCells]{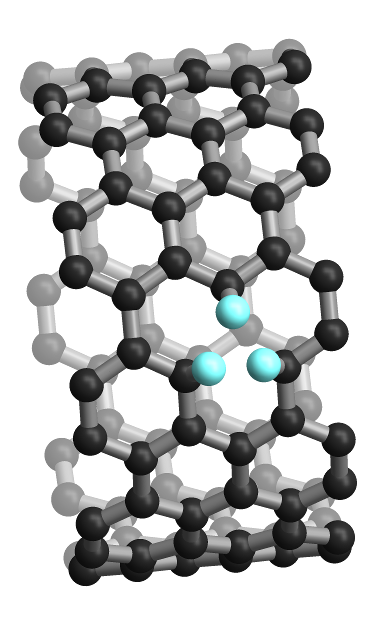}
	\end{minipage}\hfill
	\begin{minipage}{\ColumnFigure}
		\centering
		\includegraphics[width=\FigureThreeCells]{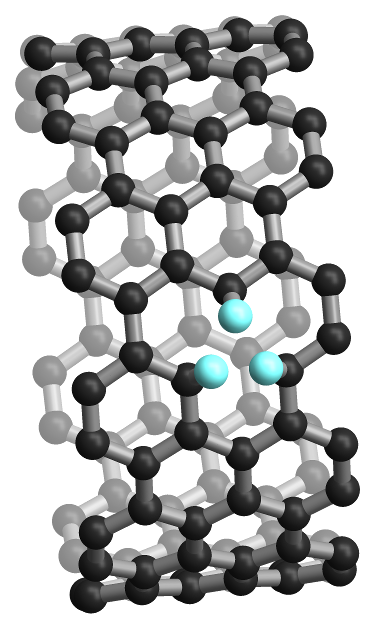}
	\end{minipage}\\
	\begin{minipage}{\ColumnCaption}
		DV$_\text{perp}$
	\end{minipage}\hfill
	\begin{minipage}{\ColumnFigure}
		\centering
		\includegraphics[width=\FigureThreeCells]{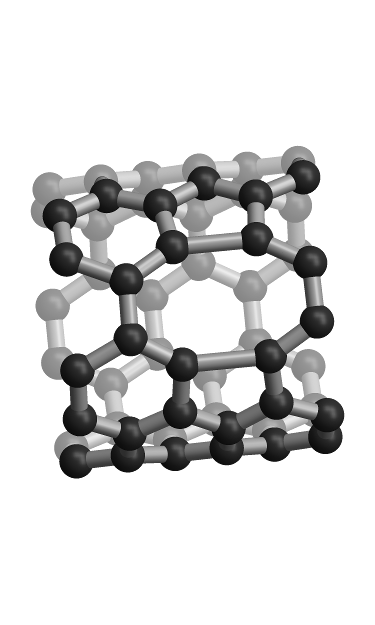}
	\end{minipage}\hfill
	\begin{minipage}{\ColumnFigure}
		\centering
		\includegraphics[width=\FigureThreeCells]{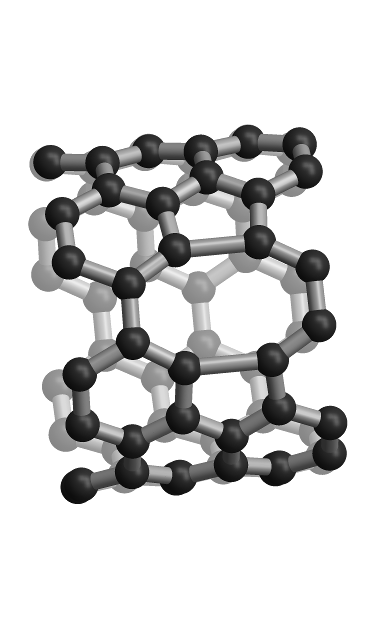}
	\end{minipage}\hfill
	\begin{minipage}{\ColumnFigure}
		\centering
		\includegraphics[width=\FigureThreeCells]{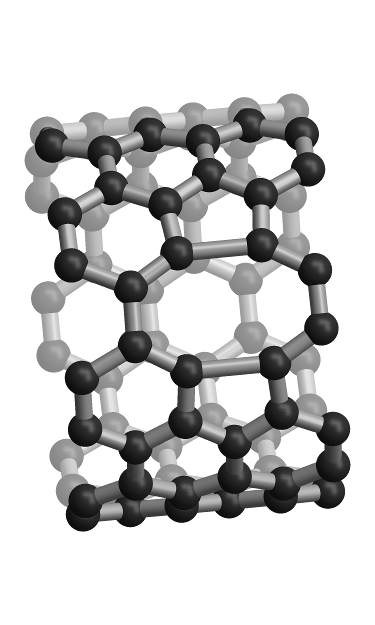}
	\end{minipage}\hfill
	\begin{minipage}{\ColumnFigure}
		\centering
		\includegraphics[width=\FigureThreeCells]{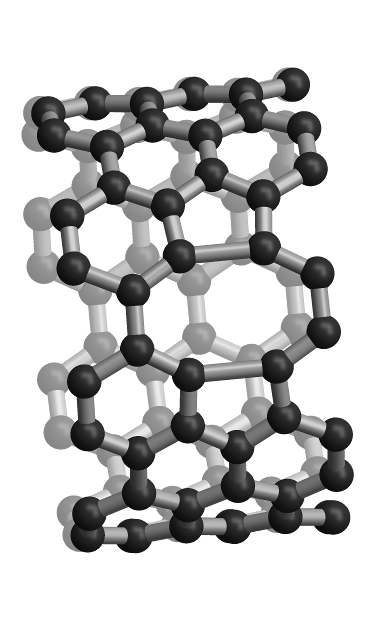}
	\end{minipage}\hfill
	\begin{minipage}{\ColumnFigure}
		\centering
		\includegraphics[width=\FigureThreeCells]{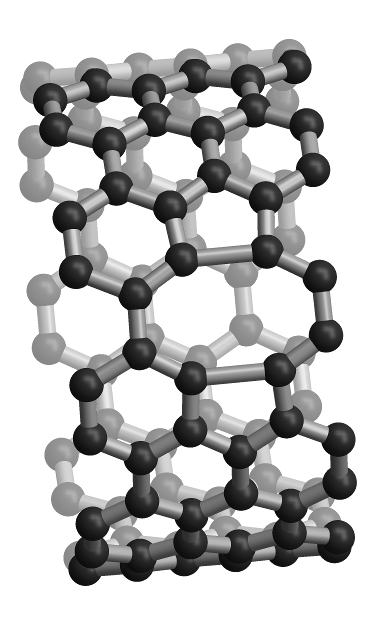}
	\end{minipage}\hfill
	\begin{minipage}{\ColumnFigure}
		\centering
		\includegraphics[width=\FigureThreeCells]{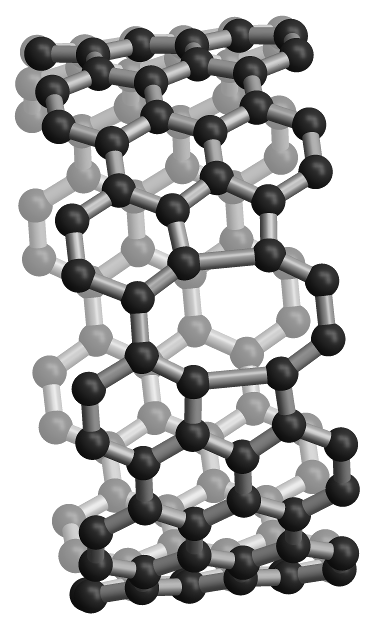}
	\end{minipage}\\
	\begin{minipage}{\ColumnCaption}
		DV$_\text{diag}$
	\end{minipage}\hfill
	\begin{minipage}{\ColumnFigure}
		\centering
		\includegraphics[width=\FigureThreeCells]{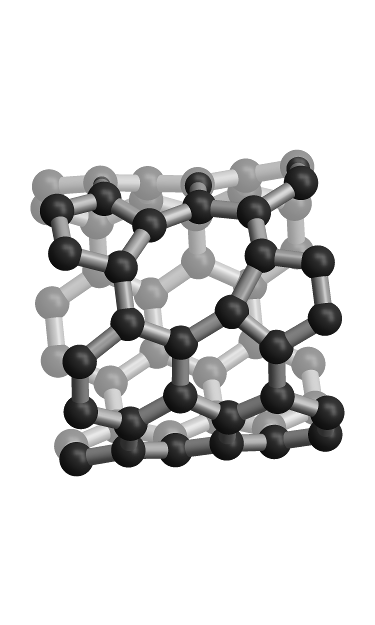}
	\end{minipage}\hfill
	\begin{minipage}{\ColumnFigure}
		\centering
		\includegraphics[width=\FigureThreeCells]{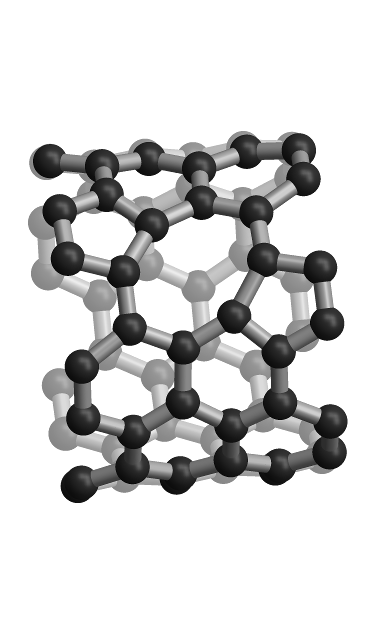}
	\end{minipage}\hfill
	\begin{minipage}{\ColumnFigure}
		\centering
		\includegraphics[width=\FigureThreeCells]{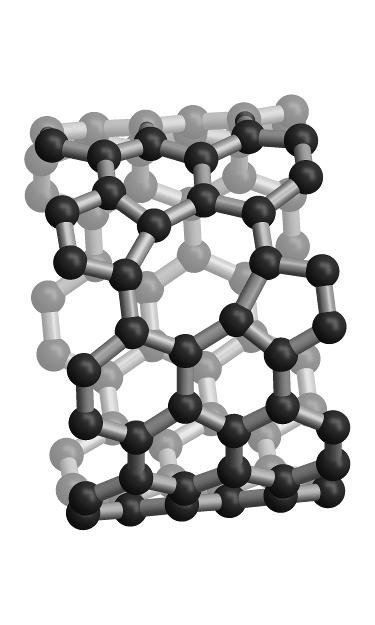}
	\end{minipage}\hfill
	\begin{minipage}{\ColumnFigure}
		\centering
		\includegraphics[width=\FigureThreeCells]{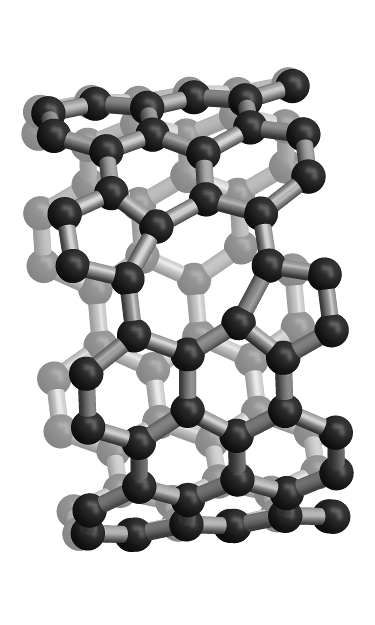}
	\end{minipage}\hfill
	\begin{minipage}{\ColumnFigure}
		\centering
		\includegraphics[width=\FigureThreeCells]{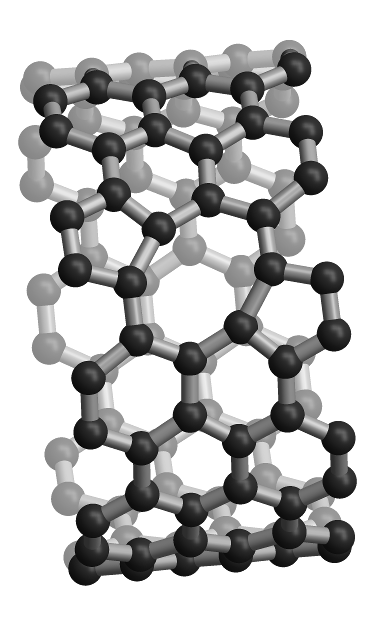}
	\end{minipage}\hfill
	\begin{minipage}{\ColumnFigure}
		\centering
		\includegraphics[width=\FigureThreeCells]{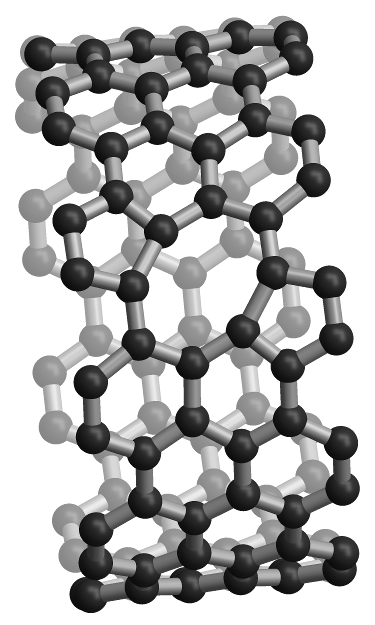}
	\end{minipage}\\
	\begin{minipage}{\ColumnCaption}
		DV$_\text{diag}$
	\end{minipage}\hfill
	\begin{minipage}{\ColumnFigure}
		\centering
		\includegraphics[width=\FigureFourCells]{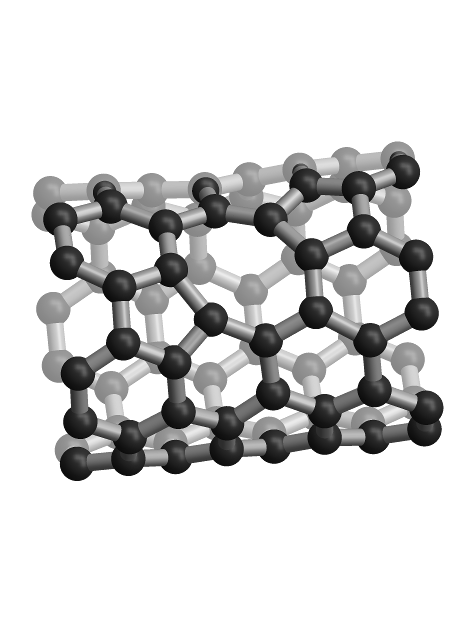}
	\end{minipage}\hfill
	\begin{minipage}{\ColumnFigure}
		\centering
		\includegraphics[width=\FigureFourCells]{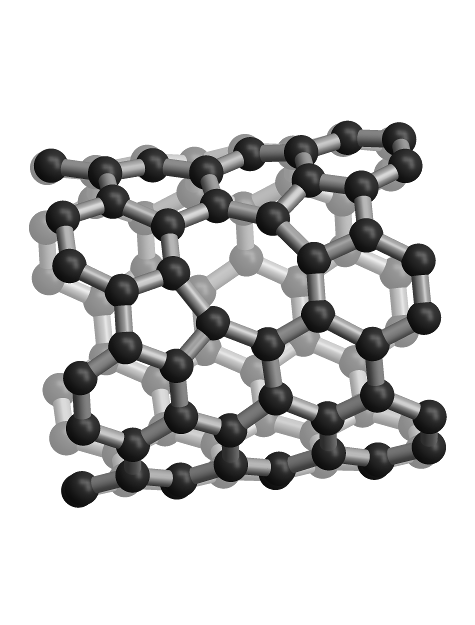}
	\end{minipage}\hfill
	\begin{minipage}{\ColumnFigure}
		\centering
		\includegraphics[width=\FigureFourCells]{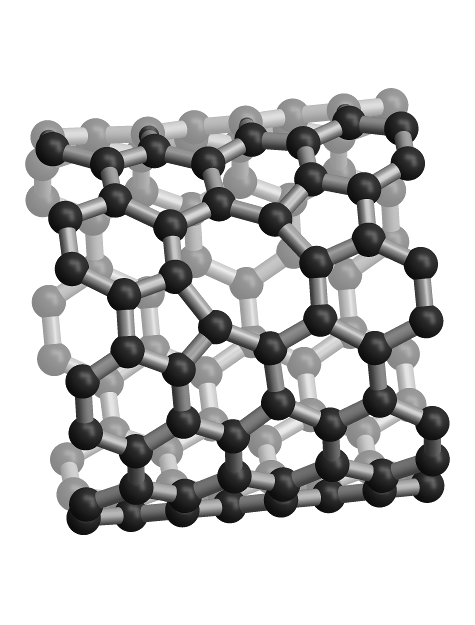}
	\end{minipage}\hfill
	\begin{minipage}{\ColumnFigure}
		\centering
		\includegraphics[width=\FigureFourCells]{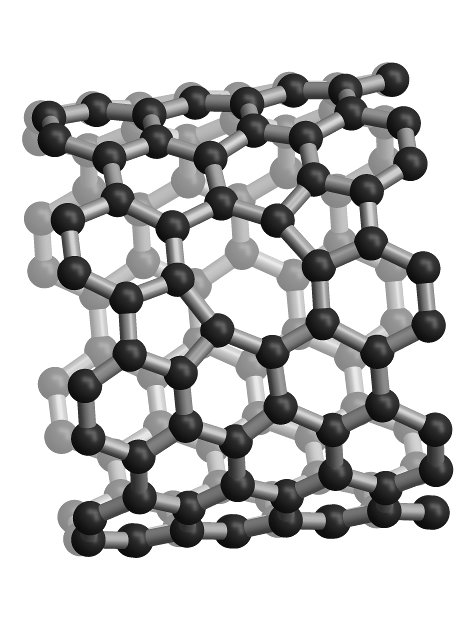}
	\end{minipage}\hfill
	\begin{minipage}{\ColumnFigure}
		\centering
		\includegraphics[width=\FigureFourCells]{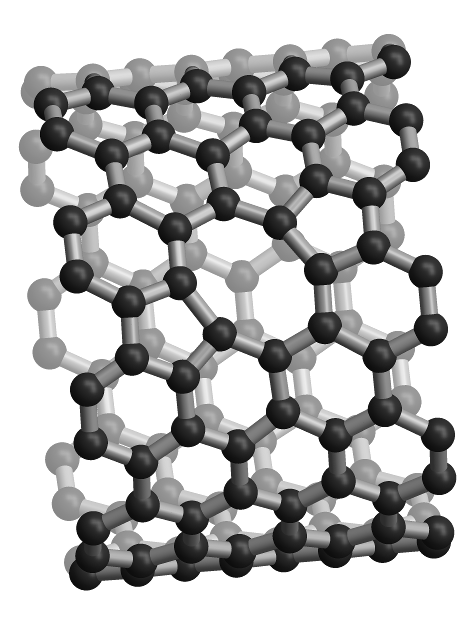}
	\end{minipage}\hfill
	\begin{minipage}{\ColumnFigure}
		\centering
		\includegraphics[width=\FigureFourCells]{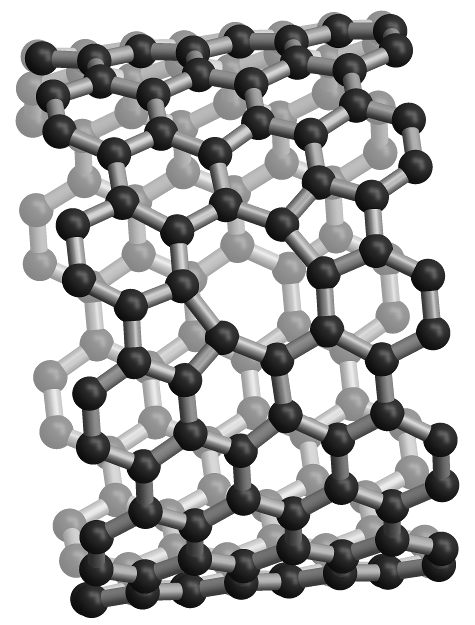}
	\end{minipage}
	\caption[Geometric structures of the CNT cells]{Geometry of the used cells for the (5,5)-CNT up to the (10,10)-CNT: unit cell of the unsaturated monovacancy (MV), saturated monovacancy (MV$_\text{3H}$), divacancy with perpendicular orientation (DV$_\text{perp}$), and the two types of the divacancy with diagonal orientation (DV$_\text{diag}$). A geometry optimization was performed for the hydrogen atoms of the MV$_\text{3H}$ defect and the whole cell of the DV defect.}
	\label{NJP:fig:defect:geometry:supp}
\end{figure}

\begin{figure}[htb]
	\includegraphics{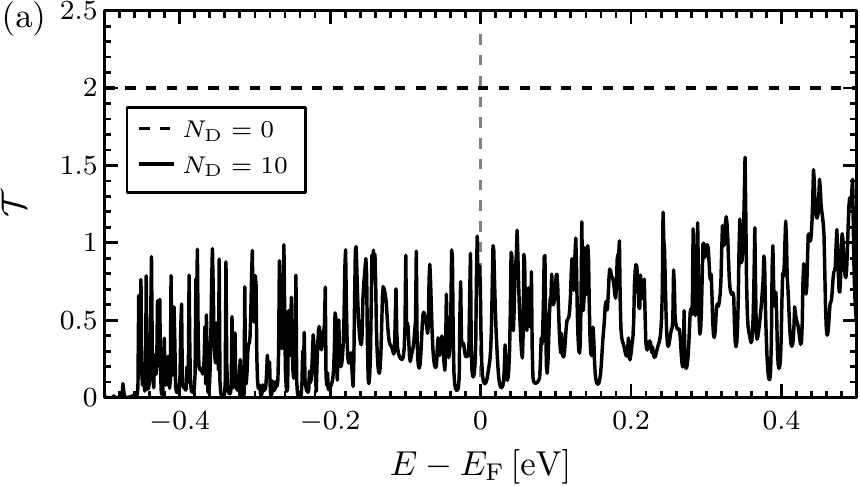}\hfill
	\includegraphics{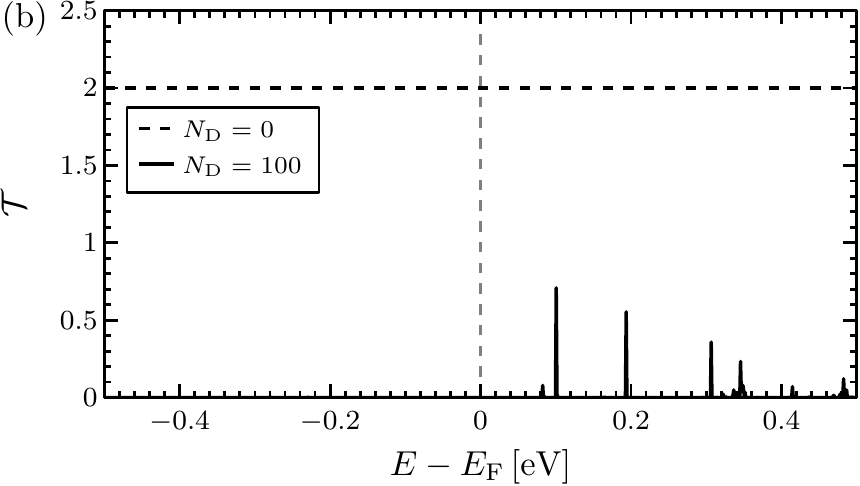}
	\caption[Transmission spectrum of a (4,4)-CNT with random defects]{Transmission spectrum of an individual (4,4)-CNT with 10 (a) and 100 (b) randomly distributed MV defects within \num{1000} cells.}
\end{figure}

\begin{figure}[htb]
	\centering
	\includegraphics{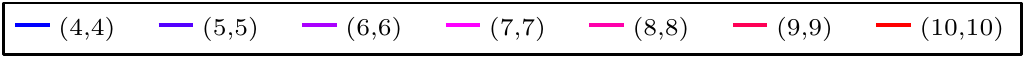}\\[0.5em]
	\includegraphics{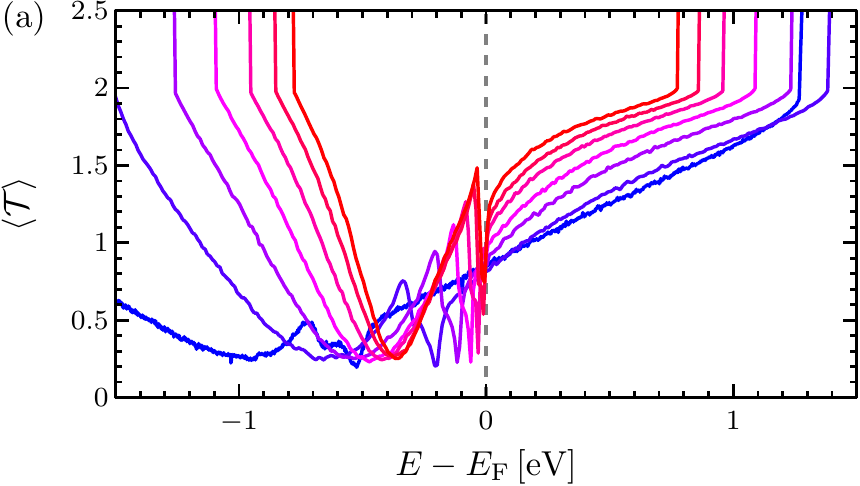}\hfill
	\includegraphics{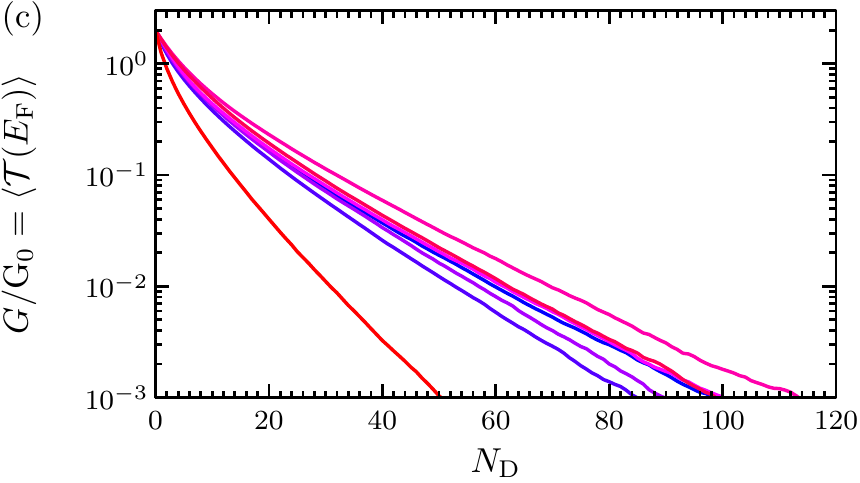}\\[0.5em]
	\includegraphics{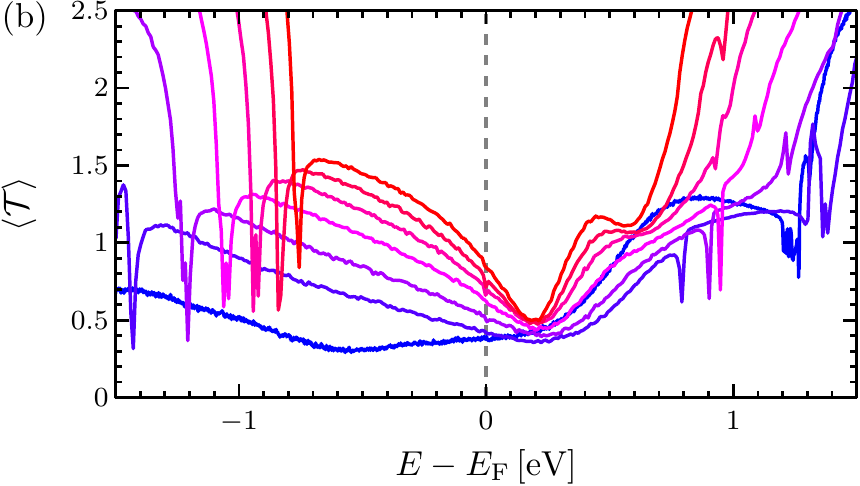}\hfill
	\includegraphics{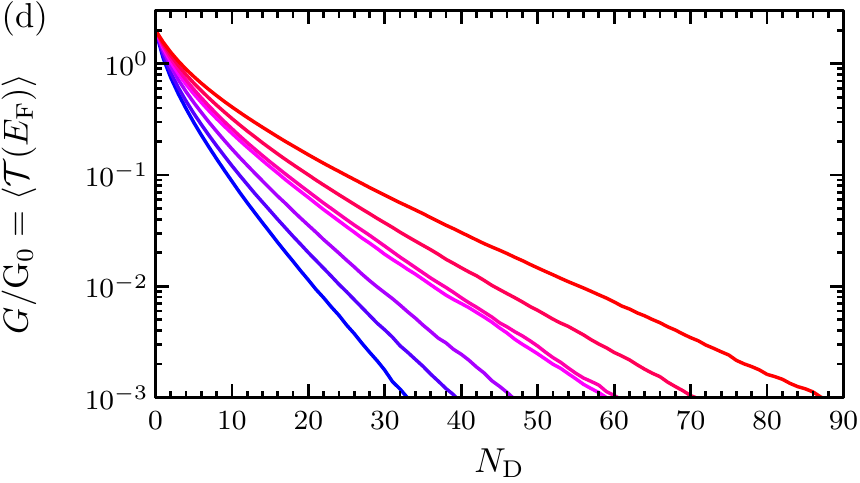}
	\caption[Transmission spectra and conductance as a function of the number of defects]{Average transmission spectra (1000 configurations, \num{1000} cells, $p_\text{D}\approx\num{0.005}$) of randomly distributed MV defects (a) and DV defects (b) for different CNTs and conductance as a function of the number MV defects (c) and DV defects (d) for different CNTs.}
\end{figure}

\begin{figure}[htb]
	\begin{minipage}{\OneColumnWidth}
		\centering
		\includegraphics{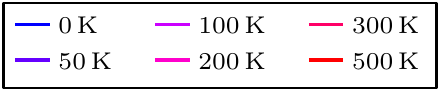}
	\end{minipage}\hfill
	\begin{minipage}{\OneColumnWidth}
		\centering
		\includegraphics{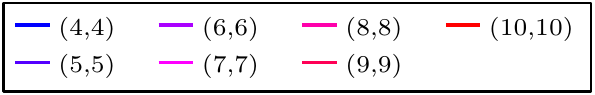}
	\end{minipage}\\[0.5em]
	\includegraphics{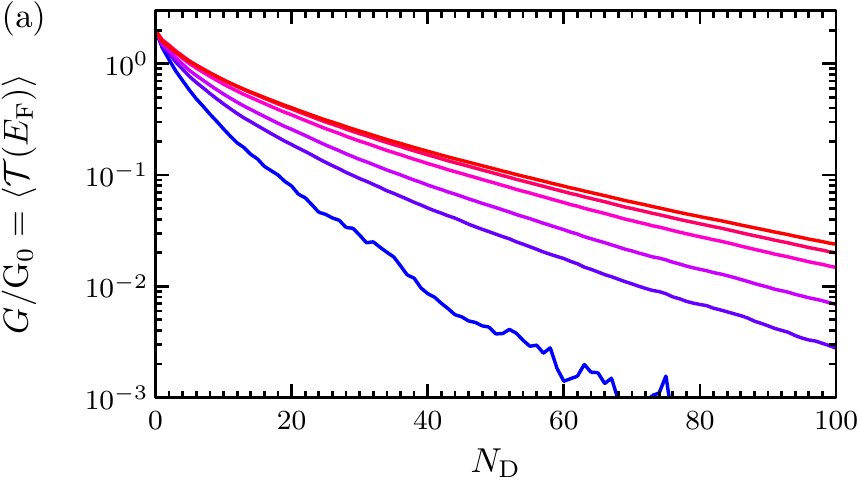}\hfill
	\includegraphics{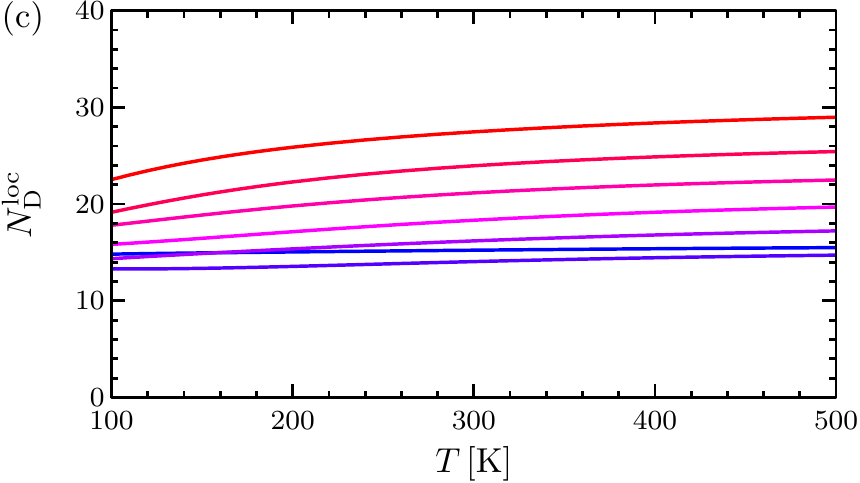}\\[0.5em]
	\includegraphics{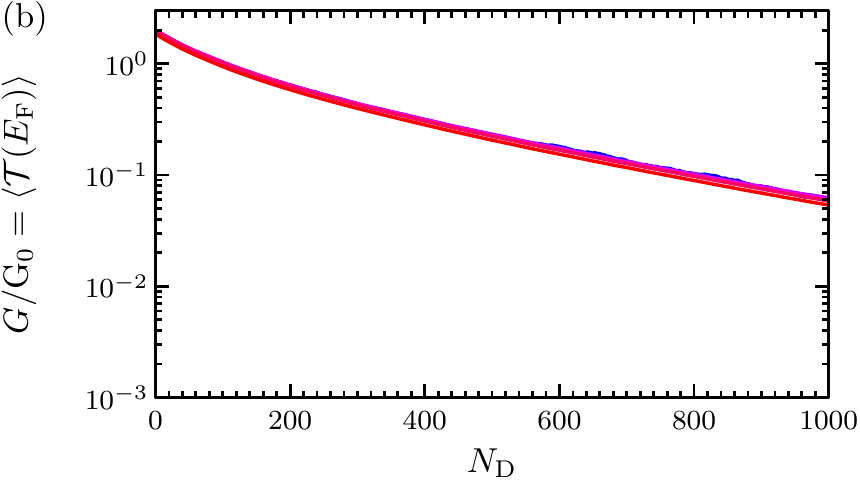}\hfill
	\includegraphics{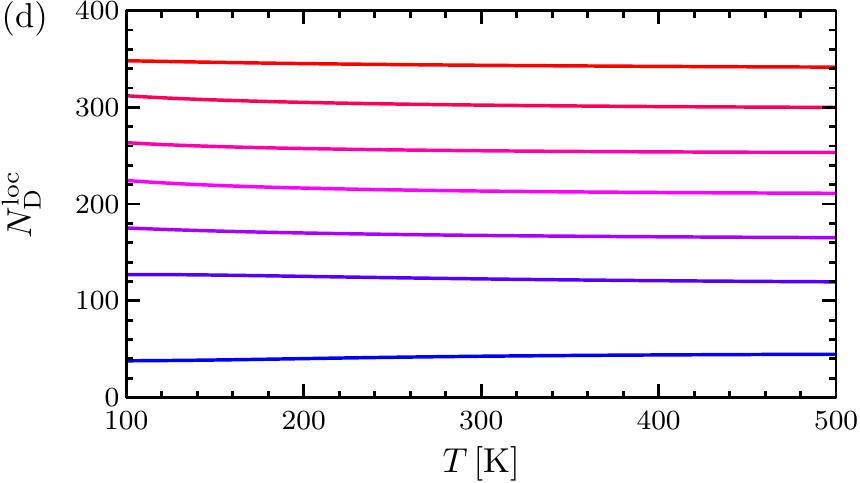}
	\caption[Conductance and localization exponent as a function of temperature]{Conductance (average over 1000 configurations) of a (10,10)-CNT in dependence on the number of MV defects ($p_\text{D}\approx\num{0.1}$, a) and MV$_\text{3H}$ defects ($p_\text{D}\approx\num{0.083}$, b) at different temperatures and localization exponent in dependence on the temperature for different CNTs with MV defects (c) and MV$_\text{3H}$ defects (d).}
\end{figure}

\begin{figure}[htb]
	\includegraphics{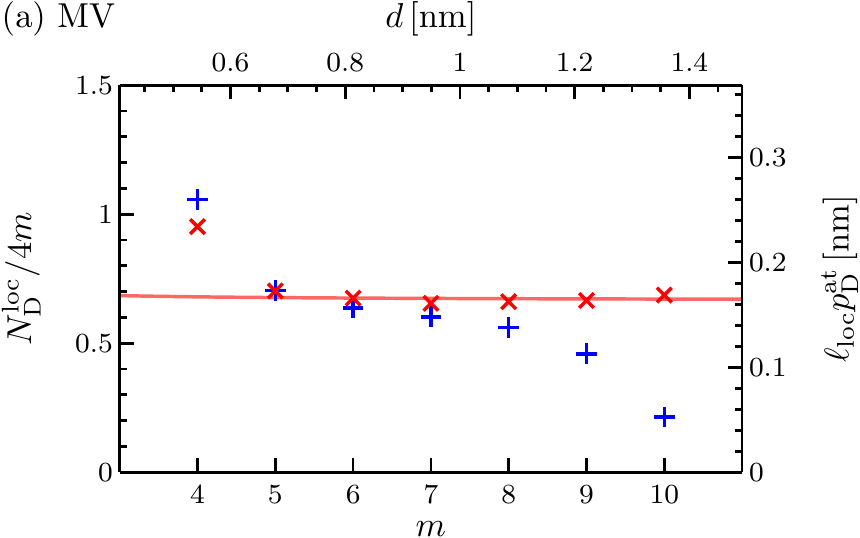}\hfill
	\includegraphics{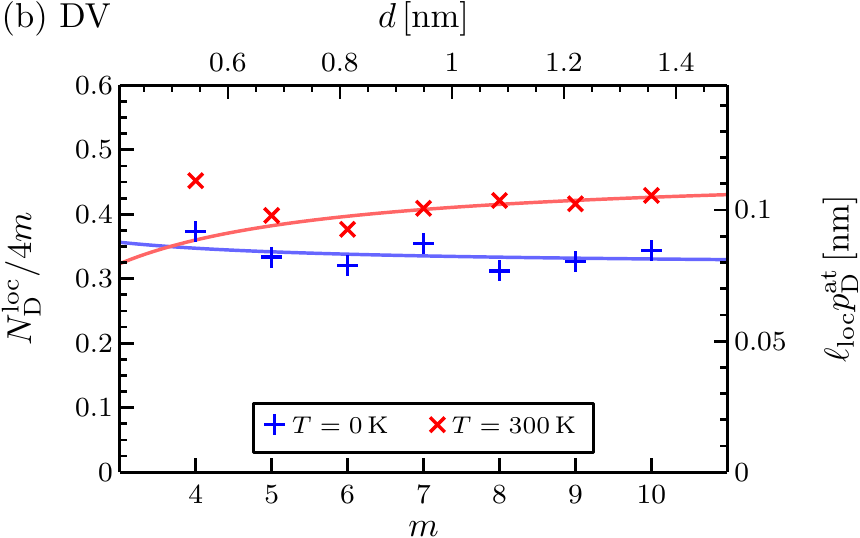}
	\caption[Localization length for constant defect probability per atom]{Diameter dependence of the localization length for constant defect probability per atom $p_\text{D}^\text{at}$ and CNTs with MV (a) and DV defects (b).  The results are shown at $T=\SI{0}{\kelvin}$ (blue, $+$) and at $T=\SI{300}{\kelvin}$ (red, $\times$). The solid lines are regressions in the range $m=5\ldots 10$ (see text for details).}
\end{figure}

\end{document}